\documentclass[%
 reprint,
 superscriptaddress,
 amsmath,amssymb,
 aps,
 showkeys,
 prb,
]{revtex4-2}

\usepackage{graphicx}
\usepackage{dcolumn}
\usepackage{bm}

\begin{document}

\preprint{APS/123-QED}

\title{Effect of a Micro-scale Dislocation Pileup on the Atomic-Scale Multi-variant Phase Transformation and Twinning}
\thanks{A footnote to the article title}%

\author{Yipeng Peng}
\affiliation{Department of Aerospace Engineering, Iowa State University, Ames, IA 50011, USA}
\author{Rigelesaiyin Ji}
\affiliation{Department of Aerospace Engineering, Iowa State University, Ames, IA 50011, USA}
\author{Thanh Phan}
\affiliation{Department of Aerospace Engineering, Iowa State University, Ames, IA 50011, USA}
\author{Laurent Capolungo}
\affiliation{Materials Science and Technology Division, Los Alamos National Laboratory, Los Alamos, NM 87545, USA}
\author{Valery~I. Levitas}
\email{vlevitas@iastate.edu}
\affiliation{Department of Aerospace Engineering, Iowa State University, Ames, IA 50011, USA}
\affiliation{Department of Mechanical Engineering, Iowa State University, Ames, IA 50011, USA}
\affiliation{U.S. Department of Energy, Ames Laboratory, Ames, IA 50011, USA}
\author{Liming Xiong}
\email{lmxiong@iastate.edu}
\affiliation{Department of Aerospace Engineering, Iowa State University, Ames, IA 50011, USA}

\date{\today}

\begin{abstract}
  In this paper, we perform concurrent atomistic-continuum (CAC) simulations to:
  (i) characterize the internal stress induced by the microscale dislocation pileup at an atomically structured interface; (ii) decompose this stress into two parts, one of which is from the dislocations behind the pileup tip according to the Eshelby model and the other is from the dislocations at the pileup tip according to a super-dislocation model; and (iii) assess how such internal stresses contribute to the atomic-scale phase transformations (PTs), reverse PTs, and twinning. The main novelty of this work is to unify the atomistic description of the interface and the coarse-grained (CG) description of the lagging dislocations away from the interface within one single framework. Our major findings are: \textbf{(a)} the interface dynamically responds to a pileup by forming steps/ledges, the height of which is proportional to the number of dislocations arriving at the interface;  \textbf{(b)} the stress intensity factors are linearly proportional to the number of the dislocations in a nanoscale pileup, but upper bends to a high level when tens of dislocations are involved in a microscale pileup; \textbf{(c)} when the pre-sheared sample is compressed, a direct square-to-hexagonal PT occurs ahead of the pileup tip and eventually grows into a wedge shape. The two variants of the hexagonal phases form a twin with respect to each other; \textbf{(d)} upon a further increase of the loading, part of the newly formed hexagonal phase transforms back to the square phase. The square product phase resulting from this reverse PT forms a twin with respect to the initial square phase. All phase boundaries (PBs) and twin boundaries (TBs) are stationary and correspond to zero thermodynamic Eshelby driving forces; and \textbf{(e)} the stress intensity induced by a pileup consisting of 16 dislocations reduces the stress required for initiating a PT by a factor of 5.5, comparing with that in the sample containing no dislocations.  This work is a first characterization of the behavior of PTs/twinning resulting from the reaction between a microscale dislocation slip and an atomically structured interface. The gained knowledge will advance our understanding on how the multi-phase material behaves in many complex physical processes, such as the synthesis of multi-phase high-entropy alloys or superhard ceramics under high pressure torsion, deep mantle earthquakes in geophysics, and so on, which all involve dislocation slip, PTs, twinning, and their interactions across from the atomistic to the microscale and beyond.
\end{abstract}


\maketitle

\section{Introduction}

Dislocations, phase transformations (PTs), and twinning are three most common carriers of plastic flow and can simultaneously be activated when deforming a crystalline media \cite{masuda2021dynamical, turlo2019linear, gunkelmann2012polycrystalline, amadou2018coupling, capolungo2008nucleation, spearot2019shear,levitasScalefreeModelingCoupled2018,kumar2018role, pandey2020situ, levitas2011phase, levitas2019high}.
An understanding of these mechanisms is not only relevant to a broad range of applications, e.g., metal forming, thermomechanical treatments of materials, shape memory alloy processing, elastocaloric applications, high-pressure physics, but also widely spread in nature, e.g., in geophysical processes.

Here, a focus is placed on metals in which the viscoplastic deformation can be conditioned by the interplay between all three deformation modes: dislocation motion, martensitic PTs, and twinning.
In such scenarios, the coupling between these deformation modes will dictate the material microstructure evolution, such as PTs accompanied by the dislocation slip \cite{pengAtomisticComputationalAnalysis2019, liu2017dislocation}, and
the PTs promoted by the dislocation- or twin-mediated plastic flow \cite{zhao2017atomic, kumar2018role}.
Several attempts on the study of the complex interaction between the dislocation-, twinning-, or PT-mediated plasticity include:
(i) a simultaneous treatment of PTs, dislocations, and twinning within a continuum sharp interphase approach from the theoretical point of view in \cite{levitas1998thermomechanical,levitas2000structural, levitas2000structural2} and the numerical point of view in \cite{idesman1999elastoplastic, idesman2000structural};
(ii) a smeared continuum description of the interaction of plasticity and PTs under high pressure;\cite{levitas2010modeling, levitas2010modeling2, feng2019fem}
(iii) continuum-level computational analysis of the role of interfaces and twinning on the PTs through crystal plasticity finite element (CPFE) simulations, where twinning is found to promote the $\alpha$ to $\omega$ PTs but suppress the reverse PTs in zirconium \cite{kumar2018role};
(iv) phase field approach (PFA) consideration of discrete dislocations and PTs at the nanoscale in \cite{kundin2011phase, levitas2012advanced, levitas2013phase,levitas2015interaction, javanbakht2015interaction} and also at the microscale in \cite{levitasScalefreeModelingCoupled2018, esfahaniStraininducedMultivariantMartensitic2020}; and
(v) atomistic simulations of interaction between PTs and plasticity in \cite{gunkelmann2012polycrystalline, kadau2002microscopic, kadau2005atomistic, wang2015coupling}.
For a more comprehensive summary of the research progress in this field, a few representative review articles on this topic can be found in \cite{olson1986dislocation, lovey1999shape, chowdhury2017revisit, vsittner2018coupling} and in \cite{fischer1994continuum, levitas2019continuum, levitas2019high, levitas2021phase} from the materials and mechanics point of view, respectively.
It should be noted that, in the above literature, twinning in martensite is often considered as one part of the multi-variant martensitic PTs.

In this work, we aim to probe the mechanisms underlying the interaction between dislocation slip, PTs, reverse PTs, and twinning.
A co-operation of these three deformation modes is pertinent to materials subjected to extreme mechanical environments (e.g., high pressure, high torsion, shock loads, etc.) \cite{bridgman2013shearing, bridgman1947effect, blank2013phase, edalati2011plastic, straumal2019phase, levitas2019high}.
Especially, the occurrence of dislocations and twinning may have a potential to decrease the critical pressure required for initiating PT.
For example: (i) an irreversible PT from rhombohedral to superhard cubic boron nitride (BN) was obtained in \cite{levitas2002low} at 5.6 GPa under plastic straining but at 55 GPa under hydrostatic loading.
A plastic strain-induced PT from hexagonal to superhard wurtzitic BN can be induced under a pressure as low as 6.7GPa in a rotational diamond anvil cell (RDAC) while under hydrostatic loading it was not observed even at 52.8 GPa \cite{ji2012shear};
(ii) the plastic strain-induced PTs from graphite to hexagonal and cubic diamonds were observed in RDAC at 0.4 and 0.7 GPa, respectively, while under hydrostatic conditions they occurred at 20 GPa and 70 GPa, respectively \cite{gao2019shear}.
Thus, the PT pressure reduction induced by a plastic straining may reach up to two orders of magnitude.
Other than contributing to the PT pressure reduction, one more surprising effect is that the plastic straining under pressure may lead to new (hidden) phases that were not or could not be reached under hydrostatic conditions \cite{bridgman2013shearing, bridgman1947effect, novikov1999new, blank2013phase, levitas2012high, levitas2019high, edalati2011plastic}.

In order to understand the dislocation-PT interaction under high pressure, the fundamental difference between the plastic strain-induced PTs and the pressure- or stress-induced PTs was introduced \cite{levitas2019continuum, levitasHighpressureMechanochemistryConceptual2004, levitasHighpressureMechanochemistryConceptual2004}.
In details, the stress-induced PT refers to the PT under stresses below the yield strength.
It starts at the pre-existing defects.
In contrast, the strain-induced PT occurs at the defects generated during the deformation.
Particularly, the dislocation pileup at the grain boundaries (GBs) is considered as the strongest internal stress concentrator responsible for the strain-induced PT \cite{levitas2019continuum, levitasHighpressureMechanochemistryConceptual2004}.
As stated above, the plastic strain-induced PTs can occur at pressures one to two orders of magnitude lower than that under hydrostatic conditions.
It thus requires a completely different thermodynamic/kinetic description and experimental characterization.
The analytical solutions in \cite{levitas2019continuum, levitasHighpressureMechanochemistryConceptual2004} derived from a model by treating the dislocation pileup as a super-dislocation shows that, because all the stress components at a pileup tip are proportional to the number of the dislocations in a pileup, the PT pressure can be reduced by a factor of 10 and more.
This conclusion was elaborated within a nanoscale PFA of coupling the evolution of discrete dislocations with PTs in \cite{levitas2013phase, levitas2015interaction, javanbakht2015interaction,levitasPhaseTransformationsNanograin2014, javanbakht2016phase, javanbakht2018nanoscale} and a microscale (or scale-free) PFA simulating the simultaneous shear (dislocation or twinning) and PT in a polycrystalline aggregate under compression and shear \cite{levitasScalefreeModelingCoupled2018, esfahaniStraininducedMultivariantMartensitic2020}.

Despite the success in providing explanation for the plastic strain-induced PT promotion, the above analytical treatments and nano-/micro-scale phase field simulations have a clear limitation: the GBs at which the dislocations pile up are modeled as fixed (with respect to material) lines, which cannot resolve the actual processes of interaction between dislocations and GBs.
In order to capture more details about the slip-interface reaction and the subsequent structure changes, an atomistic resolution at the slip-interface intersection is necessary.
Otherwise, the dislocation pileup's contribution to the subsequent PTs or twinning near the interface might be either under- or over-estimated.
For example, our previous molecular dynamics (MD) simulations demonstrated that shuffle screw dislocations transmit through tilt GBs in silicon (Si) \cite{chen2019slip}.
They simply cannot pile up and thus may not contribute much to a PT in Si.
By contrast, 60$^o$ shuffle dislocations pile up at GBs and significantly promote the amorphization \cite{chen2019amorphization}.
Also, quite often the dislocations are located in the high-pressure phase and piled up against an immobile PB, causing PT within the remaining low-pressure phase.
Such a situation has not been resolved at the atomic level yet due to difficulties of avoiding the interface motion.

Clearly, it remains a challenge using single-scale techniques/methodologies to address the full complexity associated with the microscale dislocation pileup and the subsequent structure changes because:
(a) a dislocation pileup may be tens of micrometer in length and introduce a high internal stress field spanning tens of microns away from the slip-interface intersection \cite{brittonStressFieldsGeometrically2012,guoSlipBandGrain2014};
(b) the PT at the tip of a dislocation pileup compete with slip transmission, twinning \cite{Bieler2010MMT, wangTwinNucleationSlip2010, Bieler2014, kumar2018role}, crack initiation \cite{martin2012hydrogen, kacherDislocationInteractionsGrain2014, Raabe2017AM}, which all originate from the atomic scale and need to be resolved at a high resolution.
Here we present a concurrent multiscale computational framework to address this challenge with a focus on understanding how a microscale dislocation pileup interacts with an atomically structured interface and how it contributes to the subsequent direct PTs, reverse PTs, and then twinning.
It bridges the length scale gap between atomistic and continuum, and provides us with a platform for understanding the interactions between dislocation slip, PTs, and twinning from the bottom up.
\bigskip

\section{Methodology}

A concurrent atomistic-continuum (CAC) approach \cite{xiongConcurrentSchemePassing2012, xiongConcurrentAtomisticContinuum2015, xuSequentialSlipTransfer2016, xiongNucleationGrowthDislocation2012, xuMeshRefinementSchemes2016, xiongPredictionPhononProperties2014, xuPyCACConcurrentAtomisticcontinuum2018, chenSpatialDecompositionParallel2018, chenPassingWavesAtomistic2018, xuValidationConcurrentAtomisticcontinuum2017, xiongCoarsegrainedAtomisticSimulation2011, xiongCoarsegrainedSimulationsSinglecrystal2009, Chen2019JAP, xiong2021multiscale} built upon a formulation \cite{ChenLee2003PA, ChenLee2005PM, chenLocalStressHeat2006,chenReformulationMicroscopicBalance2009,chen2016a,chen2016b,chen2018} that unifies the atomistic and continuum description of materials within one framework is deployed here.
This formulation \cite{chenReformulationMicroscopicBalance2009, chen2016a, chen2016b, chen2018} is a generalization of the Irving-Kirkwood procedure \cite{kirkwoodStatisticalMechanicalTheory1946, kirkwoodStatisticalMechanicalTheory1947, irvingStatisticalMechanicalTheory1950, bearmanStatisticalMechanicsTransport1958} in statistical mechanics.
It views the solid material as a collection of lattice cells continuously distributed in space, within each of which a group of discrete atoms is embedded.
The continuum-level physical quantities, including mass density, linear momentum density, energy density, momentum flux (also referred as stress in continuum mechanics), and energy flux, are then defined in terms of the atomic positions, velocities, and interatomic forces through Dirac or Gaussian distribution functions \cite{chenLocalStressHeat2006, chenPhysicalFoundationConsistent2018}.
An introduction of these physical quantities into the classical Newtonian mechanics leads to a series of equations, i.e., mass conservation, momentum balance, and energy conservation equations, which can govern the behavior of materials by considering them as a collection of atoms \cite{chenReformulationMicroscopicBalance2009}.
These equations are partial differential equations in the same form as the balance equations in classical continuum mechanics but with atomistic information being built-in.
Thus, those equations can be solved using numerical techniques, such as finite-difference or finite element (FE), which are commonly used for solving the equations in continuum mechanics.
The CAC simulation tool is an FE implementation \cite{xiongCoarsegrainedAtomisticSimulation2011, xiongCoarsegrainedSimulationsSinglecrystal2009} of this formulation.
\bigskip

Unlike the FE model in classical continuum mechanics, which considers the material as a collection of mass points without any internal structures, the material body in CAC is discretized into finite number of elements, each of which is a collection of the lattice cells with the atomistic information (crystal structure, slip planes, cleavage planes, interstitial sites, etc.) being embedded.
Furnished with such atomistic information, comparing with the other concurrent multiscale models, CAC has several unique features:
\textbf{(i)} it does not need additional constitutive rules and can be directly driven by the traditional interatomic potentials or the machine learning-based potentials to be trained from \emph{ab initio} calculation data;
\textbf{(ii)} it does not need any special treatments to describe discontinuity, such as dislocations, PTs, or cracking, at the continuum level but with their atomistic nature being retained;
\textbf{(iii)} it also does not need any sophisticated rules for passing the dislocation-mediated plastic flow from the atomistic domain to the continuum domain and vice versa \cite{xiongConcurrentSchemePassing2012, xiongConcurrentAtomisticContinuum2015,xuSequentialSlipTransfer2016}.
At a fraction of the cost of full MD, CAC simulations have been performed to study:
\textbf{(a)} the dislocation nucleation and growth in Cu, Al, Ni, and Si \cite{xiongNucleationGrowthDislocation2012};
\textbf{(b)} the Si-I $\rightarrow$ Si-II PT in Si \cite{xiongCoarsegrainedSimulationsSinglecrystal2009}; and
\textbf{(c)} the dislocation transmission across a twin boundary (TB) in bi-crystalline metals \cite{xuSequentialSlipTransfer2016}.
It has a predictive capability at approximately the same level as that of MD but demands significantly less computational resources. This thus provides us with an ideal platform for understanding the microscale dislocation slip and the subsequent PTs/twinning nearby an atomically structured interface.

Here, as a first demonstration of its applicability in solving such problems, a two-dimensional (2D) two-phase material (the hexagonal and square phases co-exist with an incoherent interface in between) under compression and shear is selected as a model system because:
\textbf{(i)} the 2D set-up enables a direct comparison of the simulation results with the analytical solutions under plane strain conditions for validation purpose;
\textbf{(ii)} if desired, the interface structure in such systems can be manipulated to model the interfaces in a few recently developed high-performance metallic composites, such as Ti/Al \cite{Babu-1}, Mg/Nb \cite{Sid-1}, Cu/Nb \cite{Irene-1}, and among several others.
\textbf{(iii)} the dislocation-induced local stresses in many realistic 2D materials (crystalline solids consisting of a single layer of atoms), such as one single layer of graphene sheet, boron nitride \cite{Lebedeva2016PRB, Lebedeva2019PRB, Lebedeva2020PRL, cellini2021pressure}, or colloidal crystals \cite{peng2015two, Zaccarelli2018PRL}, may have significantly contributed to the PTs in them but has not been fully understood yet.

The rest of this paper is structured as follows.
In Section \ref{Computer Model Set-up}, we briefly introduce the interatomic potential, the crystal structure, the PT variants, the boundary conditions, and also the loading strategies.
The atomistic together with the multiscale computational analysis on the dislocation pileup formation process, the pileup-induced stress accumulation, as well as its role in the subsequent PTs, reverse PTs, and twinning are then presented in Section \ref{results}.
Thereafter, we conclude this paper with a summary of our major findings as well as a brief discussion of future research in Section \ref{conclusion}.
\bigskip

\begin{figure}
  \centering
  \includegraphics[width=0.4\paperwidth]{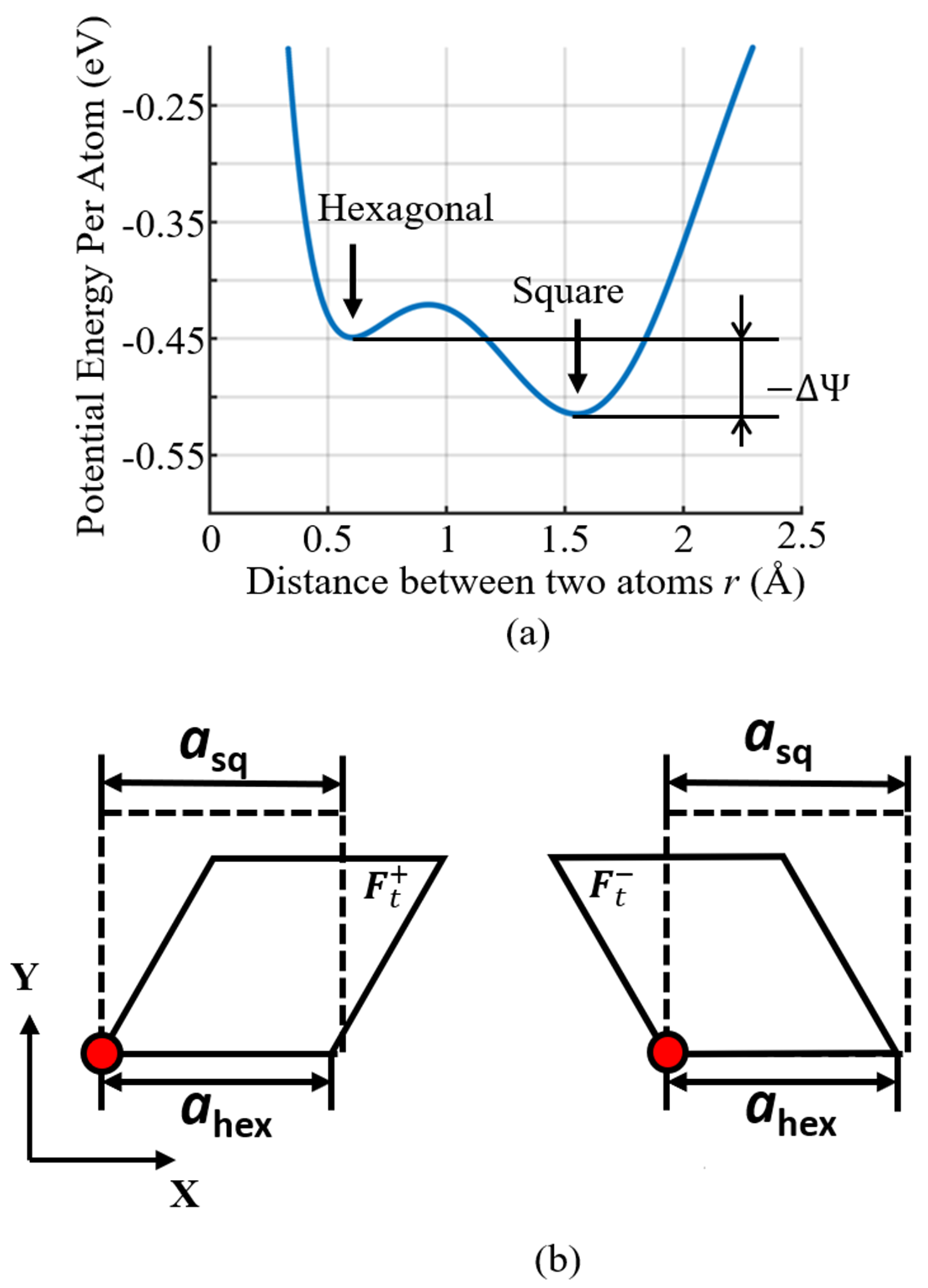}
  \caption{\label{fig:potential} (a) The potential energy surface for 2D square and hexagonal lattices with the atomic interactions in them being described by a modified Lennard-Jones potential in Eq.\ (\ref{eq:potential}); (b) two crystallographically equivalent variants ($\boldsymbol{F}_{t}^{+}$ and $\boldsymbol{F}_{t}^{-}$) of hexagonal phase (solid lines) and their orientation with respect to a square phase (dash lines).}
\end{figure}

\section{SIMULATION DETAILS}
\label{Computer Model Set-up}
\subsection{Interatomic Potential and PT Variants}

For the chosen material system, the atomic interaction in both square and hexagonal lattices is described using a modified Lennard-Jones (L-J) potential by Lee \cite{leeMechanismPressureinducedMartensitic1989} as shown in Eq.\ (\ref{eq:potential}):
\begin{equation}
  V_{MLJ}=-4\varepsilon[(\frac{\sigma}{r})^6-(\frac{\sigma}{r})^{12}]-\frac{H}{\sigma_h\sqrt{2\pi}}e^{[-\frac{(r-r_{mh})^2}{2\sigma_h^2}]},
  \label{eq:potential}
\end{equation}

\noindent which was originally proposed to study the bcc-to-hcp PT in iron.
The modified functional form of the L-J potential in Eq.\ (\ref{eq:potential}) is the combination of a traditional L-J (12-6) term and an inverse Gaussian term.
The 12-6 term leads to a stable hexagonal crystal structure at zero stress.
The addition of an inverse Gaussian into the L-J potential gives rise to a square lattice.
Here the parameters $H$, $r_{mh}$, and $\sigma_h$ in Eq.\ (\ref{eq:potential}) were chosen to stabilize:
\textbf{(i)} an incoherent interface between the square and the hexagonal phase, which acts as an obstacle to the dislocation motion; and
\textbf{(ii)} the core structure of a dislocation in the hexagonal phase.
The parameters satisfying these two conditions are listed in Table\ \ref{table:potential}.

\begin{table*}
  \caption{\label{table:potential}
    The parameters of the modified L-J potential for the square and the hexagonal phases.}
  \begin{ruledtabular}
    \begin{tabular}{ccccccc}
      mass (g/mole) & $\sigma$ (\AA{}) & $\varepsilon$ (eV) & \emph{H} (eV$\cdot$\AA{}) & $r_{mh}$ (\AA{}) & $\sigma_h$ (\AA{}) & cut-off (\AA{}) \\
      \hline
      63.546        & 2.277            & 0.415              & -0.4964                   & 3.6432           & 0.4772             & 5.0094          \\
    \end{tabular}
  \end{ruledtabular}
\end{table*}

In the absence of chemical heterogeneity, the atomic structure of materials with certain crystallographic configuration, such as fcc, bcc, hcp, diamond, zinc blend, and so on, can be determined by one single parameter: the potential energy per atom.
The calculation of the potential energy as a function of the interatomic separation, i.e., the $E-r$ relation, is often used to provide researchers with important information, such as:
(i) the $r$ at which $E$ becomes minimal is the equilibrium distance, $r_0$, between atoms in the ground state;
(ii) the pressure-induced PT from one phase to another phase occurs along the common tangent line of the $E-r$ curves of those two phases.
According to the potential energy functional form in Eq.\ (\ref{eq:potential}) and its parameters in Table\ \ref{table:potential}, the calculated $E-r$ curve is presented in Fig.\ \ref{fig:potential}a.
Three key observations from Fig.\ \ref{fig:potential}a are:
(1) this potential energy landscape has a minima at 0.595$\AA$ and 1.525$\AA$ for the hexagonal and square phases, respectively;
(2) the square phase should be the ground state because it has a lower potential energy minimum than that of the hexagonal phase; and
(3) the potential energy difference between hexagonal and square phases under zero stress is $\Delta\Psi=0.066$ eV.
Although the square-to-hexagonal PT occurs under non-zero stresses, our simulations confirm that the hexagonal phase itself is metastable and can exist under zero stress.
In this way, the chosen potential indeed enables us to stabilize two different lattices within one model.

\begin{figure*}
  \centering
  \includegraphics[width=0.8\paperwidth]{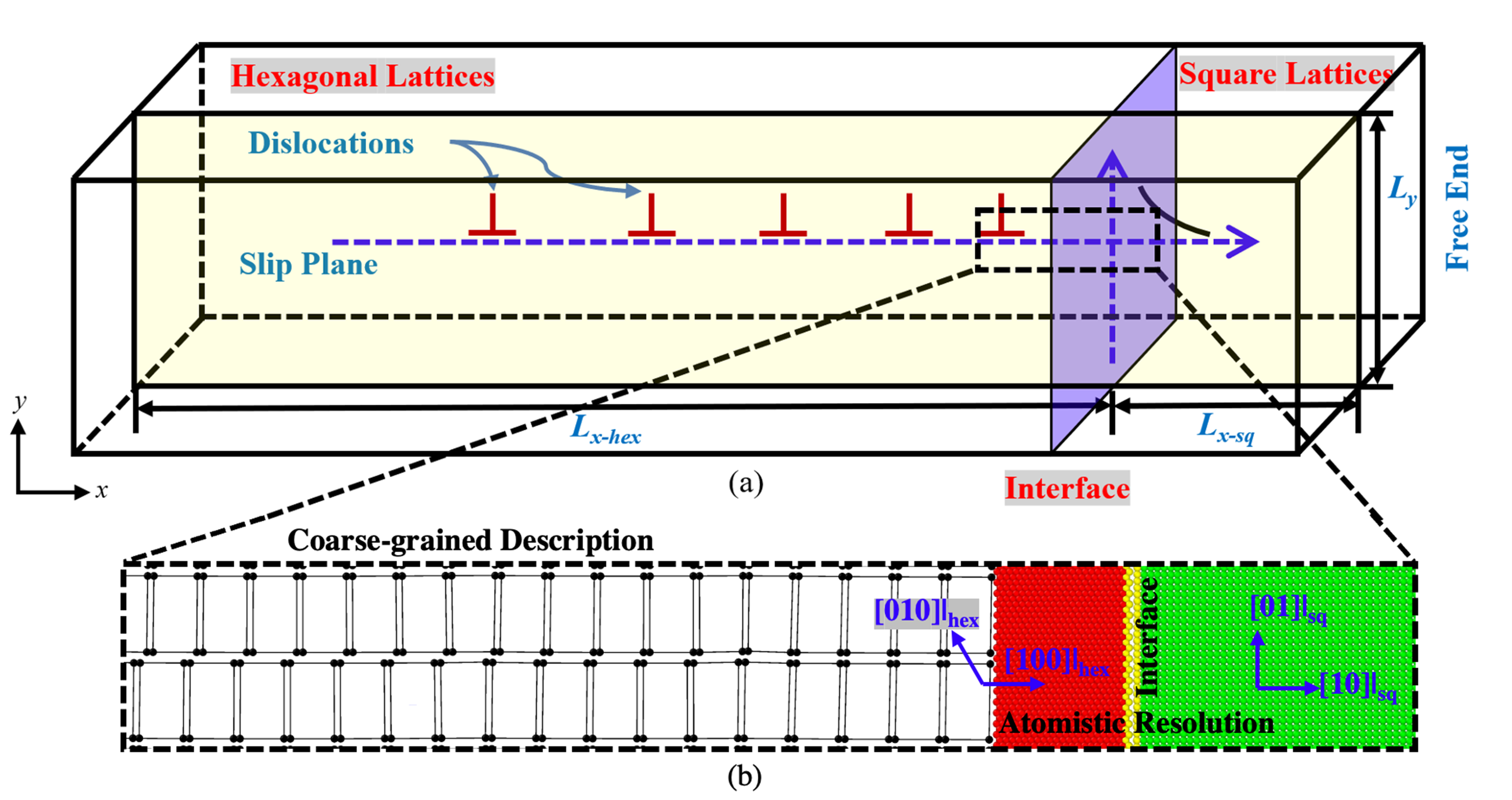}
  \caption{The CAC model setup for simulating a dislocation slip-interface interaction in a 2D two-phase material: (a) the CG description of dislocations in the hexagonal phases away from the interface and the atomistic resolution near the interface; and (b) a zoom-in display of the FE and the atomic configuration in the dashed box of (a).}
  \label{fig:model}
\end{figure*}

Figure\ \ref{fig:potential}b shows the unit cell of the square (dash lines) and the hexagonal (solid lines) phases with their lattice parameters being noted as $a_{\mathrm{sq}}$ and $a_{\mathrm{hex}}$, respectively.
In a 2D set-up, the transformation deformation gradient tensor $\boldsymbol{F}_t$, the stress and strain tensors can be all written as 2$\times$2 matrices in a cartesian system with orthogonal unit basis vectors.
Clearly, there are two crystallographically equivalent variants for the hexagonal phase due to the lattice symmetry.
These two variants are in a twinned configuration with respect to each other.
According to the lattice cell set-up of these two phases in Fig.\ \ref{fig:potential}b, the transformation deformation gradient associated with those two variants resulting from the square-to-hexagonal PTs are:
\begin{equation}
  \boldsymbol{F}_t^+= \left(\begin{matrix}  0.941 &  0.471 \\  0 & 0.815 \end{matrix}\right);\ \
  \boldsymbol{F}_t^-= \left(\begin{matrix}  0.941 &  -0.471 \\  0 & 0.815\end{matrix}\right).
  \label{eq:F}
\end{equation}
\noindent The transformation deformation gradient can be then decomposed into orthogonal and symmetric parts through $\boldsymbol{F}_t=\boldsymbol{R} \cdot \boldsymbol{U}$.
The symmetric part is:
$\boldsymbol{U}_t=\left(\boldsymbol {F}_t^T \cdot \boldsymbol{F}_t \right)^{1/2}$.
Here the superscript $T$ means transposition.
As such, the symmetric (rotation-free) transformation deformation gradient for these two variants can be calculated as
\begin{equation}
  \boldsymbol{U}_t^+= \left(\begin{matrix}  0.909 &  0.244 \\   0.244 & 0.909 \end{matrix}\right);\ \
  \boldsymbol{U}_t^-= \left(\begin{matrix}  0.909 & -0.244 \\  -0.244 & 0.909 \end{matrix}\right).
  \label{eq:U}
\end{equation}
\bigskip

\subsection{The Computer Model Setup}

In order to determine the role of a dislocation pileup in the subsequent structure changes near the interface, the CAC model for a two-phase material system with square and hexagonal lattices co-existing is constructed (Fig.\ \ref{fig:model}a).
Clearly, when this system is subjected to loading, the deformation behavior near the square/hexagonal interface is critical and may dictate the material's overall microstructure evolution.
The material domain near the interface is thus resolved at a fully atomistic resolution (Fig.\ \ref{fig:model}a).
In contrast, away from the interface, a CG description of the hexagonal phase using FE is deployed (Fig.\ \ref{fig:model}b), which has much less degrees of freedom (DOF) than that of a fully atomistic model.
To study the slip-interface reaction, similar to our previous work \cite{xiongConcurrentSchemePassing2012, xiongConcurrentAtomisticContinuum2015, xuSequentialSlipTransfer2016}, the FEs in the CG domain are carefully aligned with their boundaries along a dislocation slip plane in the hexagonal lattice.
Here, each FE contains sixty-four atoms and can slide with each other along the element boundaries.
In this way, the dislocation-mediated slip far away from the interface can be accommodated in the CG domain at a fraction of the cost of MD.
More importantly, since each FE in Fig.\ \ref{fig:model}b is a collection of lattice cells, the forces acting on the FE nodes are calculated by converting the interatomic forces into the internal force density using a Gauss quadrature scheme elucidated in \cite{xiongCoarsegrainedAtomisticSimulation2011, xiongCoarsegrainedSimulationsSinglecrystal2009}.
As such, the constitutive rule for the atomistic and CG domains in the CAC model is the same, i.e., the interatomic potential in Eq.\ (\ref{eq:potential}).
This differs from other continuum approaches which accommodate dislocations through the deployment of either a contact model in \cite{levitasScalefreeModelingCoupled2018, esfahaniStraininducedMultivariantMartensitic2020}, a Heaviside step function in \cite{gracieConcurrentlyCoupledAtomistic2009}, or an additional DOF in CPFE \cite{Hartmaier2012AM, McDowell2014JMPS}.
\bigskip

\begin{figure*}
  \centering
  \includegraphics[width=0.8\paperwidth]{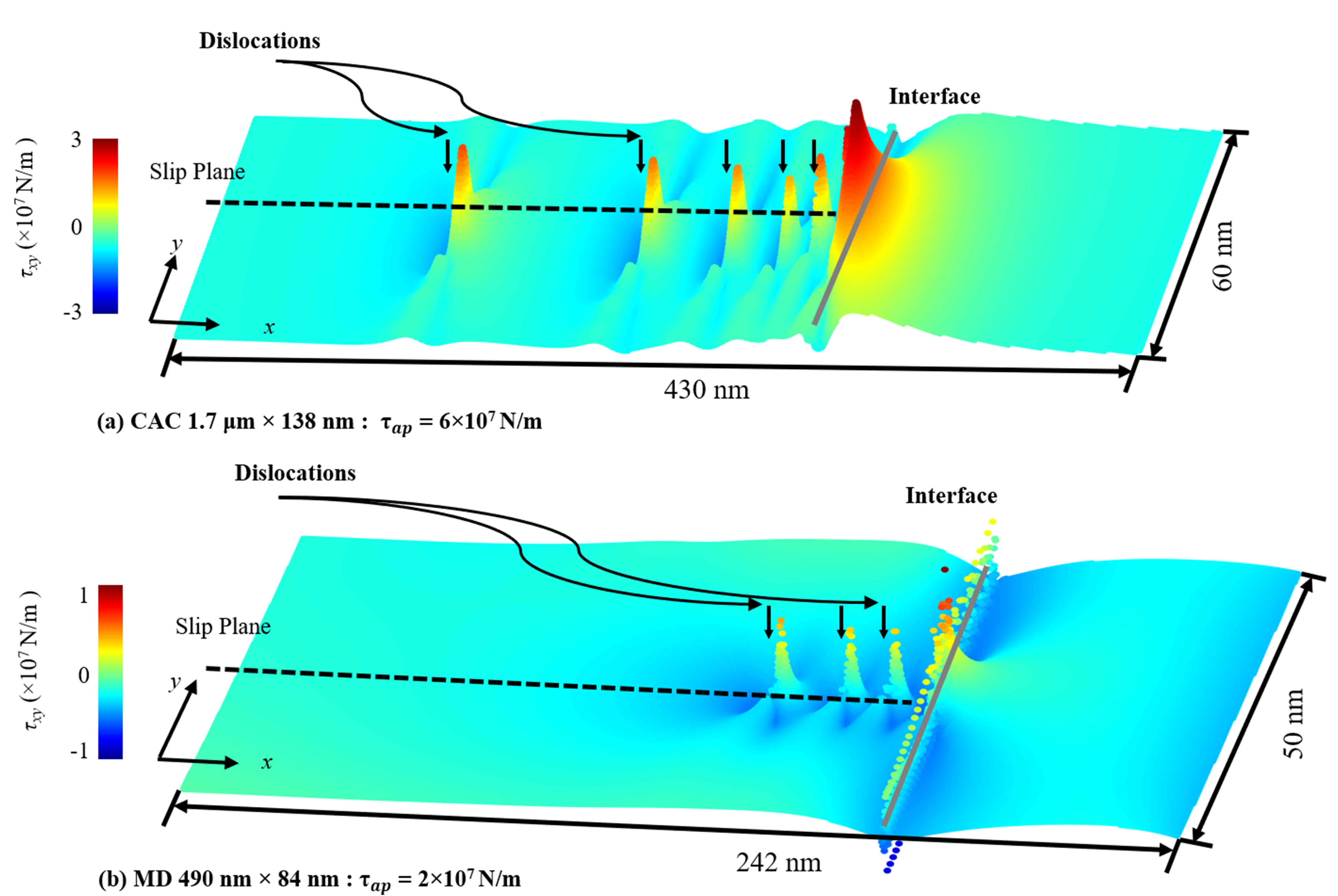}
  \caption{The dislocation pileup configuration and the contour of the internal stresses (the applied stress, $\tau_{ap}$, is not included here) from (a) microscale CAC: 16 dislocations are piled up against the interface under a shear $\tau_{ap} = 6\times10^7\ \mathrm{N/m}$; 6 of them arrive at the interface; (b) nanoscale MD: 8 dislocations are piled up at $\tau_{ap} = 2\times10^7\ \mathrm{N/m}$; 5 of them arrive at the interface.}
  \label{fig:stresscontour}
\end{figure*}

Here we have carefully chosen the crystallographic orientation of both the hexagonal and square lattices to construct the incoherent interface (Fig.\ \ref{fig:model}b) such that:
\textbf{(a)} the initially introduced dislocation slip is perpendicular to the interface, and \textbf{(b)} a dislocation transmission across the interface is suppressed, which has been justified through a detailed Schmid factor analysis and geometric compatibility factor analysis in \cite{peng2022atomistic}.
In this way, a large number of dislocations can be piled up at the interface and a high internal stress is generated ahead of the slip-interface intersection, whose contributions to the subsequent PTs can be then well quantified without the need of considering transmission, cross-slip, and among several other complexities.
Although the interface here satisfying the above two conditions may not be the one with an energy minimum, in practice, the majority of interfaces found in materials will not be at a minimum energy state anyway due to the presence of defects.

A queue of dislocations with a uniform separation of 12 nm in between are initially introduced into the hexagonal lattice, which is now in a CG description (Fig.\ \ref{fig:model}b).
As demonstrated in our previous work \cite{xiongConcurrentSchemePassing2012, xiongConcurrentAtomisticContinuum2015, xuSequentialSlipTransfer2016}, because the CG domain in a CAC model can accommodate dislocations without smearing out their atomistic natures, the introduction of a dislocation into the CG domain is realized here through displacing FE nodes according to the analytical solution for the displacement field around an edge dislocation derived from the theory of elasticity.
Different from traditional FE models in which the neighboring elements are connected by sharing the FE nodes, the FEs in CAC models across the slip plane are disconnected (See Fig.\ \ref{fig:model}b).
Thus, the displacement jump induced by a dislocation along the slip plane will be allowed.
It should be also noted that the dislocation displacement fields, strain and stresses resulting from present CAC simulations will deviate from the linear isotropic elasticity-based solution.
Instead, the actual atomic-level information, including crystal anisotropy, finite strain, nonlinear elasticity, and the interface structure heterogeneity will be all considered.
The reason that we initially introduce dislocations through displacing the FE nodes according to the elasticity-based solution is to accelerate the convergence of our calculations by reducing the time required for equilibrating the system.

Due to the deployment of a CG description in the region away from the interface, the hexagonal lattice dimension, noted as $L_{x-\mathrm{hex}}$ in Fig.\ \ref{fig:model}a, in the CAC model can be at the micrometer level and even above.
It enables us to initially introduce tens of dislocations into the model.
This goes beyond the reach of a traditional MD model because MD usually has a limited length scale at nanometers, which can only accommodate a few dislocations in the pileup \cite{wangAtomisticSimulationsDislocation2015}.
In the present simulations, on the other side of interface ahead of the pileup tip, the crystal structure is in a square lattice with a dimension of $L_{x-\mathrm{sq}}$ along \emph{x} direction.
The sample dimension along \emph{y} direction is chosen as $L_y=m_{\mathrm{hex}}a_{\mathrm{hex}}=n_{\mathrm{sq}}a_{\mathrm{sq}}$, where $m_{\mathrm{hex}}$ and $n_{\mathrm{sq}}$ are the numbers of the hexagonal and square lattice cells along \emph{y} direction, respectively.
In addition to microscale CAC models, a series of nanoscale MD simulations are also performed using LAMMPS \cite{plimptonFastParallelAlgorithms1995}.
In all these simulations, the displacement along \emph{z} direction of the sample is constrained to achieve a plane strain condition.
The initial crystallographic orientations of square and hexagonal lattices are indicated in Fig.\ \ref{fig:model}b.
The dimensions and the number of the degree-of-freedom (DOF) in both nanoscale MD and microscale CAC models are listed in Table\ \ref{table:dimension}.
The left and right ends of the sample along $x$ direction are free.
A homogeneous shear within the $xy$ plane is imposed until a desired shear stress is achieved.
Thereafter, the top and bottom ends of the sample along $y$ direction are fixed for equilibrating the dislocation configuration in the pileup.
When a PT needs to be activated, a displacement controlled compressive loading along $y$ direction will be then applied on the two ends of the sample.

\begin{table*}
  \caption{\label{table:dimension}
    The dimensions and the number of the DOF in MD and CAC models}
  \begin{ruledtabular}
    \begin{tabular}{cccccc}
                     & $L_{x-\mathrm{hex}}$ & $L_{x-\mathrm{sq}}$ & $L_y$  & DOF                              & Number of dislocations in a pileup \\
      \hline
      Microscale CAC & 1.58 $\mu$m          & 120 nm              & 138 nm & 315,528 FE nodes + 118,392 atoms & 16                                 \\
      Nanoscale CAC  & 420 nm               & 70 nm               & 84 nm  & 55,416 FE nodes + 20,982 atom    & 8                                  \\
      Nanoscale MD   & 420 nm               & 70 nm               & 84 nm  & 1,355,284 atoms                  & 8
    \end{tabular}
  \end{ruledtabular}
\end{table*}

\section{Simulation Results}
\label{results}
\subsection{Dislocation Pileup Process}

The computational models containing the built-in dislocations are firstly relaxed for a duration of 20 ps with a time step of 1 fs to achieve a stable configuration.
Thereafter, a shear stress, noted as $\tau_{ap}$, is imposed on the whole samples to drive the dislocations towards the interface.
A certain level of $\tau_{ap}$ is realized through applying a homogeneous shear strain while monitoring the resulting shear stress on the fly. If the shear stress does not match the desired $\tau_{ap}$, the shear stress is adjusted by increasing or decreasing the applied shear strain.
When $\tau_{ap}$ arrives at the desired value, the top and bottom boundaries of the sample are constrained and not allowed to move along both \emph{x} and \emph{y} directions any more.
\bigskip

An internal stress concentration, noted as $\tau_{xy}$, ahead of the slip-interface intersection is generated due to the large number of dislocations' arrival at the interface.
Fig.\ \ref{fig:stresscontour}a and Fig.\ \ref{fig:stresscontour}b show the contour of the internal stress field obtained from CAC and MD simulations of 16 and 8 dislocations piling up at the interface under a shear of $\tau_{ap} = 6\times10^7\ \mathrm{N/m}$ and $\tau_{ap} = 2\times10^7\ \mathrm{N/m}$, respectively.
It should be noted that the contour here shows the magnitude of $\tau_{xy}-\tau_{ap}$, rather than the absolute value of $\tau_{xy}$. It is seen that the level of the internal stress concentration ahead of a pileup tip in Fig.\ \ref{fig:stresscontour}a is significantly higher than that in Fig.\ \ref{fig:stresscontour}b, because more dislocations have participated in the formation of a pileup in CAC (Fig.\ \ref{fig:stresscontour}a) than that in MD (Fig.\ \ref{fig:stresscontour}b).

\begin{figure}
  \centering
  \includegraphics[width=0.4\paperwidth]{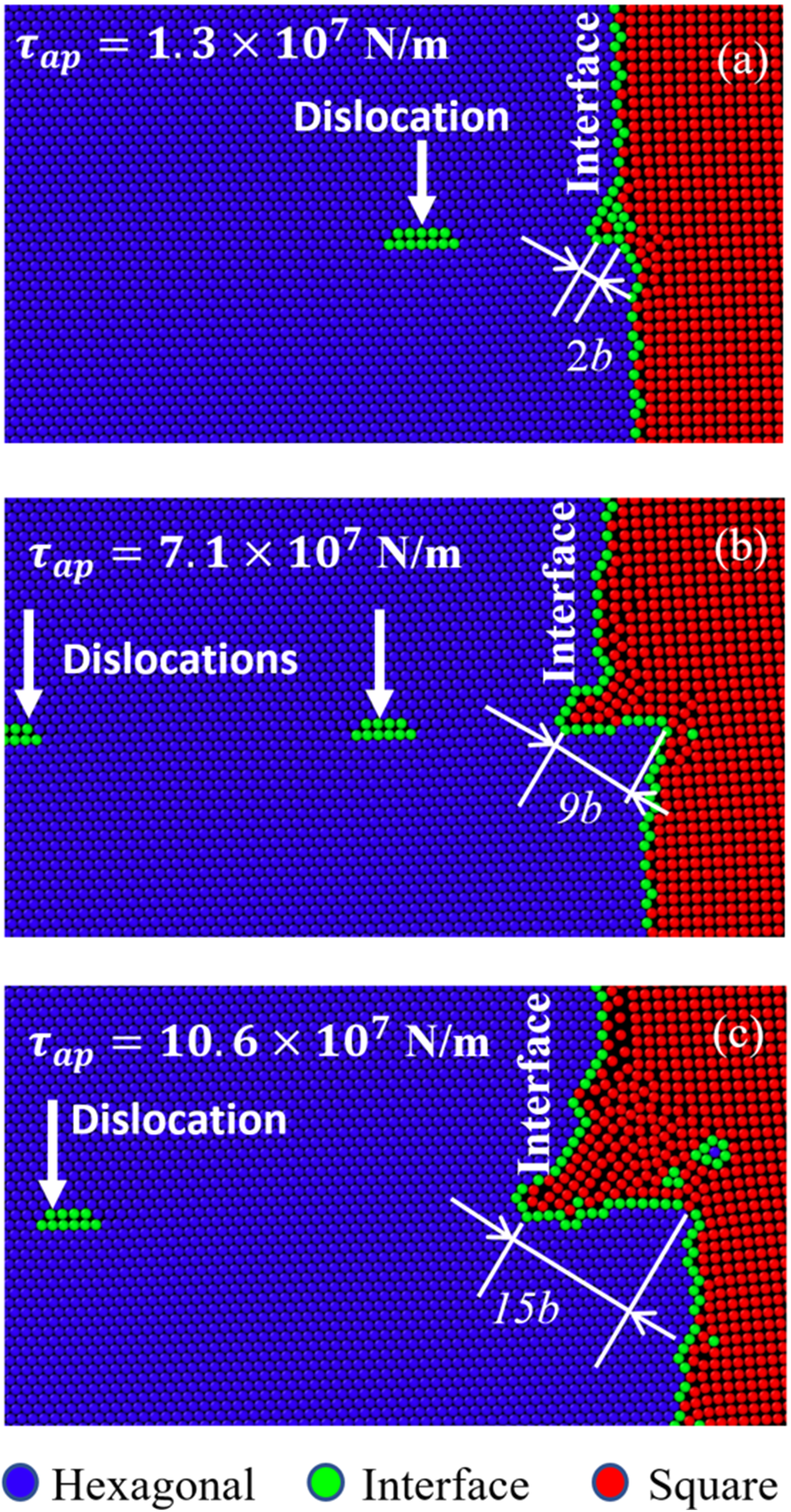}
  \caption{The atomic-scale structure evolution at the slip-interface intersection when different number of dislocations arrive at the interface under different applied stresses, $\tau_{ap}$. The atoms are color coded through an atomic-level coordination number analysis. The step formation at the slip-interface intersection is similar to a super-dislocation acting as a strong stress concentrator.}
  \label{fig:pileuptip}
\end{figure}

With an atomistic resolution at the interface, CAC provides us with an opportunity of examining the atomic-level structure evolution at the slip-interface intersection.
Figure\ \ref{fig:pileuptip} presents the snapshots showing the atomic structure configuration (red: square lattice; blue: hexagonal lattice; green: defects) when different number of dislocations arrive at the interface.
It is seen that, when $\tau_{ap}$ is at a level of $1.3\times10^7\ \mathrm{N/m}$, one dislocation is behind the pileup tip and two dislocations arrive at the interface, a step with a height of 5.12 \AA, i.e., the magnitude of two Burgers vector, $2\boldsymbol{b}$, is formed (Fig.\ \ref{fig:pileuptip}a).
This step can be approximately considered as a single super-dislocation with a Burgers vector of $N_a\boldsymbol{b}$, where $N_a$ is the number of the dislocations arriving at the interface.
Upon a further increase up to $\tau_{ap}=7.1\times10^7\ \mathrm{N/m}$, 9 dislocations arrive at the interface and eventually, leading to the formation of a step with a height of 9$b$ (Fig.\ \ref{fig:pileuptip}b).
Under a shear stress of $\tau_{ap} = 10.6\times10^7\ \mathrm{N/m}$, 15 dislocations arrive at the interface, Fig.\ \ref{fig:pileuptip}c show the corresponding atomic arrangements.
At this stage, approximately, the step at the interface has a height of around 15$b$.

As designed, the interface under consideration here indeed blocks the motion of dislocations without allowing any transmission.
Also, as shown in \cite{peng2022atomistic}, the dislocation configuration behind the pileup tip in Fig.\ \ref{fig:stresscontour} actually matches an analytical solution from Hirth and Lothe very well.
More importantly, the step-like super-dislocation with a complex local structure is formed at the slip-interface intersection.
Due to the local atomic structure reconstruction, this super-dislocation carries a sophisticated core structure rather than a simple additive of multiple dislocation cores.
Similar features were observed in our previous full MD simulations of dislocation pileup against a tilt grain boundary in silicon, in which the amorphization has been found ahead of a slip-interface intersection \cite{chen2019amorphization}.
It should be also noted that the presence of a thermal bath would likely promote the re-arrangement of the atoms at the interface and may promote a dislocation transmission.
Without including any thermal fluctuations, the present simulations should be considered as extreme cases for the slip-interface reaction.
A resolution of the thermal-induced atomic structure reconfiguration during the slip-interface reaction at finite temperature is beyond the scope here and will be studied in our future work.

\subsection{Pileup-Induced Stress Intensity Factors}

\begin{figure}
  \centering
  \includegraphics[width=0.4\paperwidth]{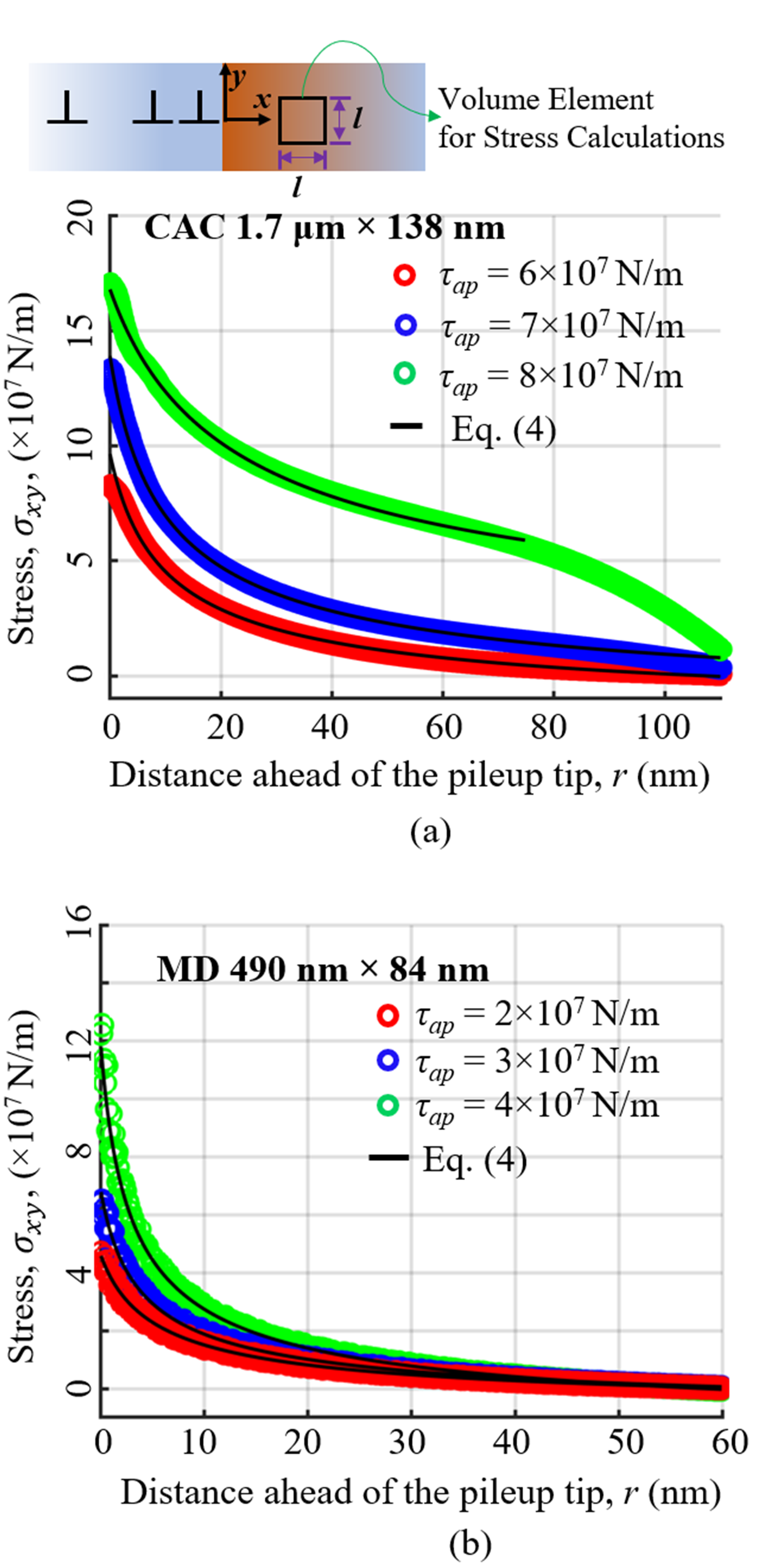}
  \caption{The local stress profile ahead of the slip-interface intersection under a variety of $\tau_{ap}$ from (a) CAC and (b) MD simulations and their fits into Eq.\ (\ref{eq:stressdistribution})}
  \label{fig:stressdistribution}
\end{figure}

To characterize the stress profile ahead of the pileup tip, we construct a series of finite-sized volume elements ahead of the slip-interface intersection (see the inset picture in Fig.\ \ref{fig:stressdistribution}).
Each volume element is at a resolution of 5 \AA $\times$ 5 \AA.
The stress associated with each volume element can be then calculated using the Virial formula.
Figure\ \ref{fig:stressdistribution} presents the CAC- and MD simulation-predicted shear stress distributions ahead of the dislocation pileup tip.
Two major observations are: (a) both CAC and MD simulations predict an obvious stress concentration at the pileup tip and its rise upon an increase of the applied shear, $\tau_{ap}$; (b) this stress concentration decays away from pileup tip but spans a longer range (100nm, Fig.\ \ref{fig:stressdistribution}a) in CAC than that (60nm, Fig.\ \ref{fig:stressdistribution}b) in MD because more dislocations have been piled up in the CAC model.
It suggests that, similar to what has been observed in experiments \cite{brittonStressFieldsGeometrically2012}, the participation of a large number of dislocations in a slip may produce a long-range internal stress field, which can span tens of microns away from the pileup tip if tens of dislocations are piled up.

\begin{figure*}
  \centering
  \includegraphics[width=0.8\paperwidth]{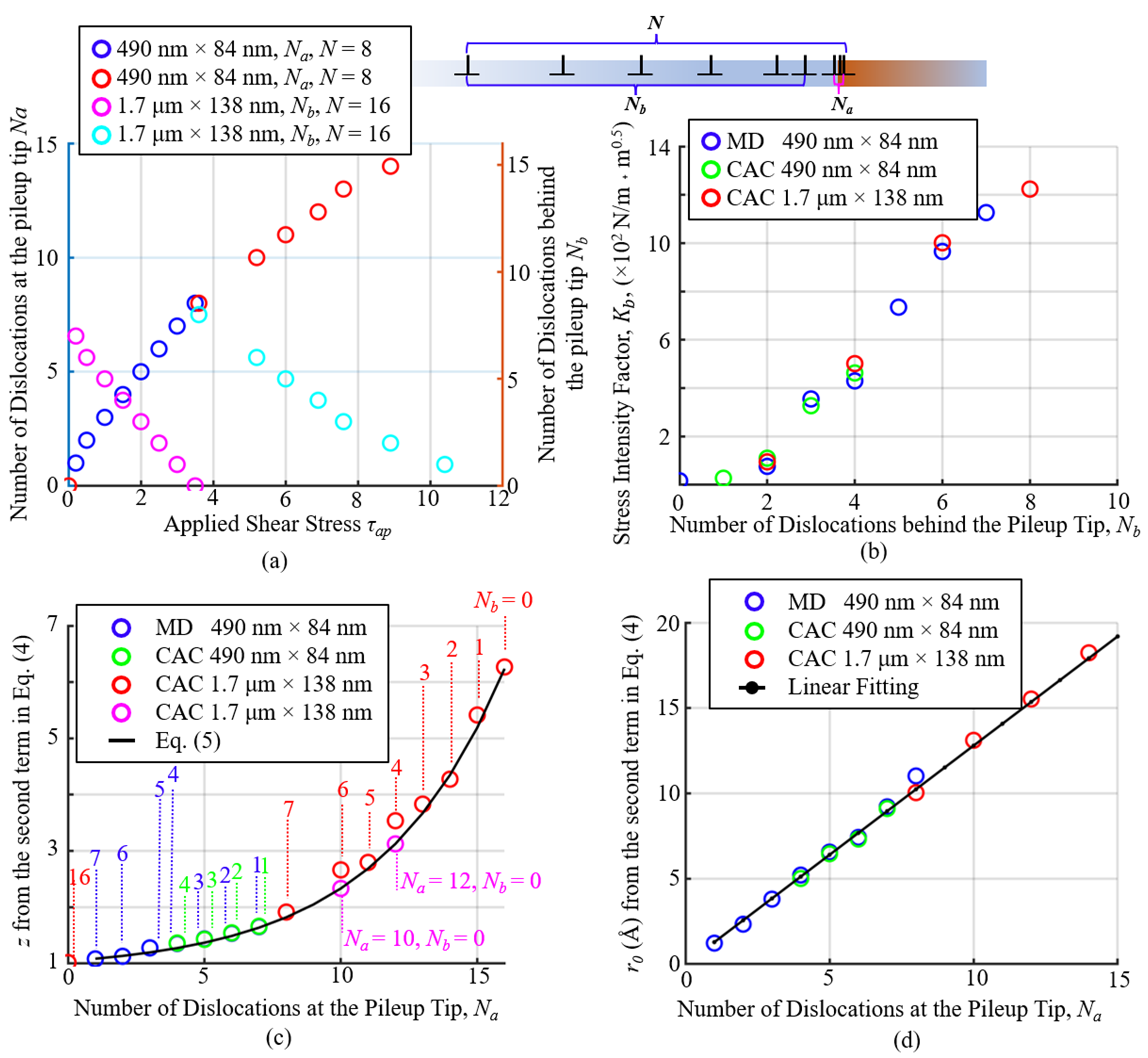}
  \caption{The nanoscale MD and microscale CAC simulation-predicted relationship: (a) the number of dislocations behind the pileup tip $N_b$, the number of dislocations at the pileup tip $N_a$ versus the applied shear stress $\tau_{ap}$; (b) between the internal stress intensity factor, $K_b$, and the number of the dislocations, $N_b$, behind a dislocation pileup tip; (c) between the correction coefficient $z$ and the number of the dislocations at the dislocation pileup tip $N_a$; (d) between the location of the maximum internal stress, $r_0$, and the number of the dislocations, $N_a$, at the dislocation pileup tip.}
  \label{fig:stressprofilefitting}
\end{figure*}

The obtained stress profile ahead of the pileup tip can be then fit into an equation as follows:

\begin{equation}
  \tau_{xy}=\frac{K_b}{\sqrt{\pi(r+r_0)}}+\frac{z \mu N_ab}{2\pi (1-v)(r+r_0)}+\tau_0.
  \label{eq:stressdistribution}
\end{equation}

\noindent In this equation, the dislocation pileup-induced internal stresses is, for the first time, decomposed into two parts.
The first term on the right side of Eq.\ (\ref{eq:stressdistribution}) is based on the Eshelby model for the dislocations behind the pileup tip which considers the interface as a rigid obstacle to dislocation motion.
The second term on the right side of Eq.\ (\ref{eq:stressdistribution}) is caused by the step, also noted as the super-dislocation at the pileup tip.
In Eq.\ (\ref{eq:stressdistribution}), $\mu=35.4\times10^7$ N/m and is the shear modulus of the materials (the hexagonal lattice in this work); $b$ is the magnitude of the Burgers vector 2.56 \AA; $v=0.3$ and is the Poisson's ratio; $N_a$ is the number of dislocations that participate the step formation at the pileup tip; $\tau_0$ is a parameter for considering the uncertainty associated with the reference stress state when no dislocation is introduced in the model; $r_0$ is the fitting parameter due to the structural change during the step formation, the value of which suggests the location of the maximum internal stress at the pileup tip; \emph{r} is the distance between the pileup tip and the stress measurement site; $K_b$ is the stress intensity factor induced by the dislocations behind the pileup tip; $z$ is the correction coefficient due to finite strain, nonlinear anisotropic elasticity, finite step size, and two-phase material.

A determination of those parameters is accomplished through fitting Eq.\ (\ref{eq:stressdistribution}) into the stress profiles from two sets of simulations: (i) in simulation set-1, for a sample containing $N$ dislocations, we apply a stress until all dislocations arrive at the interface and form a step, i.e., $N_a=N$ and $N_b=0$.
Here, $N_b$ stands for the number of the dislocations behind the pileup tip (see inset picture of Fig. 6 for the physical meaning of $N$, $N_a$, and $N_b$). When $N_a=N$ and $N_b=0$, the first term on the right side of Eq.\ (\ref{eq:stressdistribution}) simply goes to zero.
A fitting of simulation-based stress profile into Eq.\ (\ref{eq:stressdistribution}) leads to a determination of the value of $z$,  $\tau_0$, and $r_0$; (ii) in simulation set-2, a stress lower than that in set-1 is imposed on the same sample containing $N$ dislocations.
In this situation, $N_b\neq0$ and $N_a=N-N_b$.
As such, the first term on the right side of Eq.\ (\ref{eq:stressdistribution}) is non-zero.
The $K_b$ in it can be then determined by fitting the stress profile from simulation set-2 into Eq.\ (\ref{eq:stressdistribution}).
It should be noted that, at this step, the values of $z$,  $\tau_0$, and $r_0$ fitted from simulation set-1 are used.
In the fitting process, we only consider the stress profile up to $r=67$nm where the internal stress does not relax due to the presence of free surfaces.
Two different sample sizes are considered here.
One is of a dimension of 490 nm $\times$ 84 nm and contains 8 dislocations.
The other is of a dimension of 1.7 $\mu$m $\times$138 nm and contains 16 dislocations.

\begin{figure*}
  \centering
  \includegraphics[width=0.8\paperwidth]{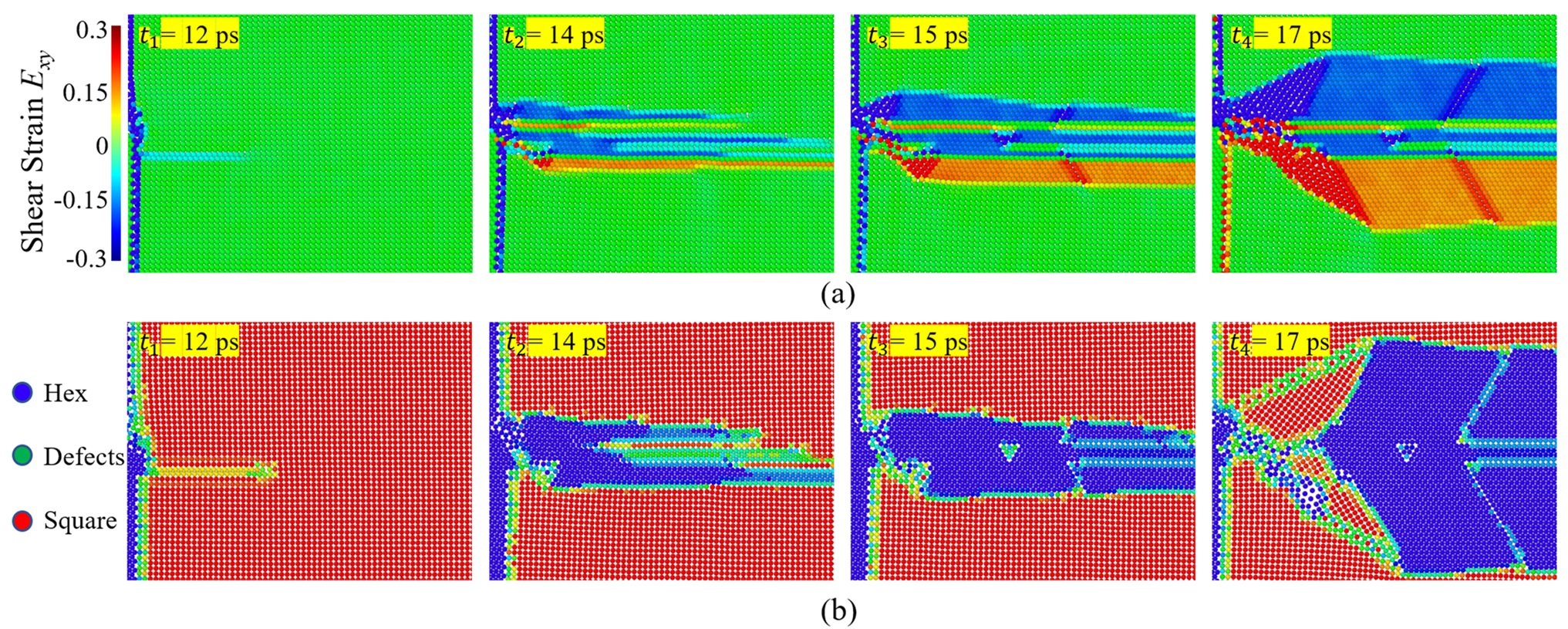}
  \caption{The time sequences of the atomic-scale structure evolution ahead of the pileup tip in CAC simulations showing the process of a square-to-hexagonal PT (direct) and then the hexagonal-to-square PT (reverse) in pre-sheared two-phase materials under compression. The atoms in the first row of this figure are colored in a shear strain of $E_{xy}$ and the second row is colored according to their coordination number with a cutoff of 3.92 \AA}
  \label{fig:grain_evo}
\end{figure*}

The $\tau_{ap}-N_a$ and $\tau_{ap}-N_b$ relations obtained from the MD together with CAC simulation of the dislocation pileup process are presented in Fig.\ \ref{fig:stressprofilefitting}a. It is seen that, in all simulations, when a pileup happens upon an increase of $\tau_{ap}$, $N_a$ increases and $N_b$ decreases. Fig.\ \ref{fig:stressprofilefitting}b presents the simulation-predicted relationship between $K_b$ and $N_b$. Clearly, $K_b$ linearly increases upon an increase of  $N_b$. In addition to $K_b$, another parameter that we have introduced into Eq.\ (\ref{eq:stressdistribution}) is the correction coefficient $z$, which is needed here because:
(1) the classical super-dislocation model only considers the defect embedded in an isotropic medium with a constant shear modulus of $\mu$ while we have considered the super-dislocation at a PB in a two-phase anisotropic material;
(2) during the dislocation pileup process, both physical and geometric nonlinearities develop;
(3) additivity of stress field from each dislocation is violated.
Especially, when more than 8 dislocations arrive at the PB, the slope of $z-N_a$ relation is significantly larger than that when $N_a$ is less than 8 (Fig.\ \ref{fig:stressprofilefitting}c).
This in turn, implies that the stress intensity factor upper bends to a very high level when tens of dislocations participate the formation of a pileup at the microscale, comparing with that when only several dislocations are involved in the formation of a pileup at the nanoscale.
Such a a highly non-linear $z-N_a$ relation can be fit into an equation as:

\begin{equation}
  z=0.1656e^{0.2173N_a}+0.8777.
  \label{eq:zfitting}
\end{equation}

It should be noted that the data in Fig.\ \ref{fig:stressprofilefitting}c exhibits a trend of deviating from Eq.\ (\ref{eq:zfitting}) when $N_b$ is larger than 2, especially when $N=16$, $N_a=12$, $N_b=4$ or when $N=16$, $N_a=10$, $N_b=6$.
We believe this deviation is caused by a fundamental difference between a super-dislocation model and the computer set-up for non-zero $N_b$.
A super-dislocation model, i.e., the second term on the right side of Eq.\ (\ref{eq:stressdistribution}), is only exact when $N_b=0$, i.e, all the dislocations arriving at the interface participate the formation of a step.
When $N_b$ is larger than 2, one can't rely on the super-dislocation model to estimate the internal stresses ahead of a slip-interface intersection any more.
To confirm this assertion, we perform two additional CAC simulations.
In one simulation, $N=12$, $N_a=12$, and $N_b=0$. By contrast, in another simulation, $N=10$, $N_a=10$, and $N_b=0$.
Comparing with the data from simulations with $N=16$, $N_a=12$, and $N_b=4$ or  $N=16$, $N_a=10$, and $N_b=6$, the results (pink circles) from these two new sets of simulations with $N_b=0$ obviously fits into the Eq. (\ref{eq:zfitting}) with a significantly less error.
Other than $z$ and $K_b$, another key parameter in Eq.\ (\ref{eq:stressdistribution}) is $r_0$, which indicates the location of the maximum stress concentration.
This location obviously changes during the pileup process.
$r_0$ is thus introduced in Eq.\ (\ref{eq:stressdistribution}) to capture the such local structure evolution at the pileup tip.
As expected, the values of $r_0$ obtained from fitting simulation data into Eq.\ (\ref{eq:stressdistribution}) are found to be linearly proportional to the number of the dislocations arriving at the interface (Fig.\ \ref{fig:stressprofilefitting}d) with the slope equals to $b/2$, $b$ as the length of the Burgers vector.
This means that $r_0$ is located in the middle of a step produced by dislocations at the interface.
\bigskip

\begin{figure*}
  \centering
  \includegraphics[width=0.8\paperwidth]{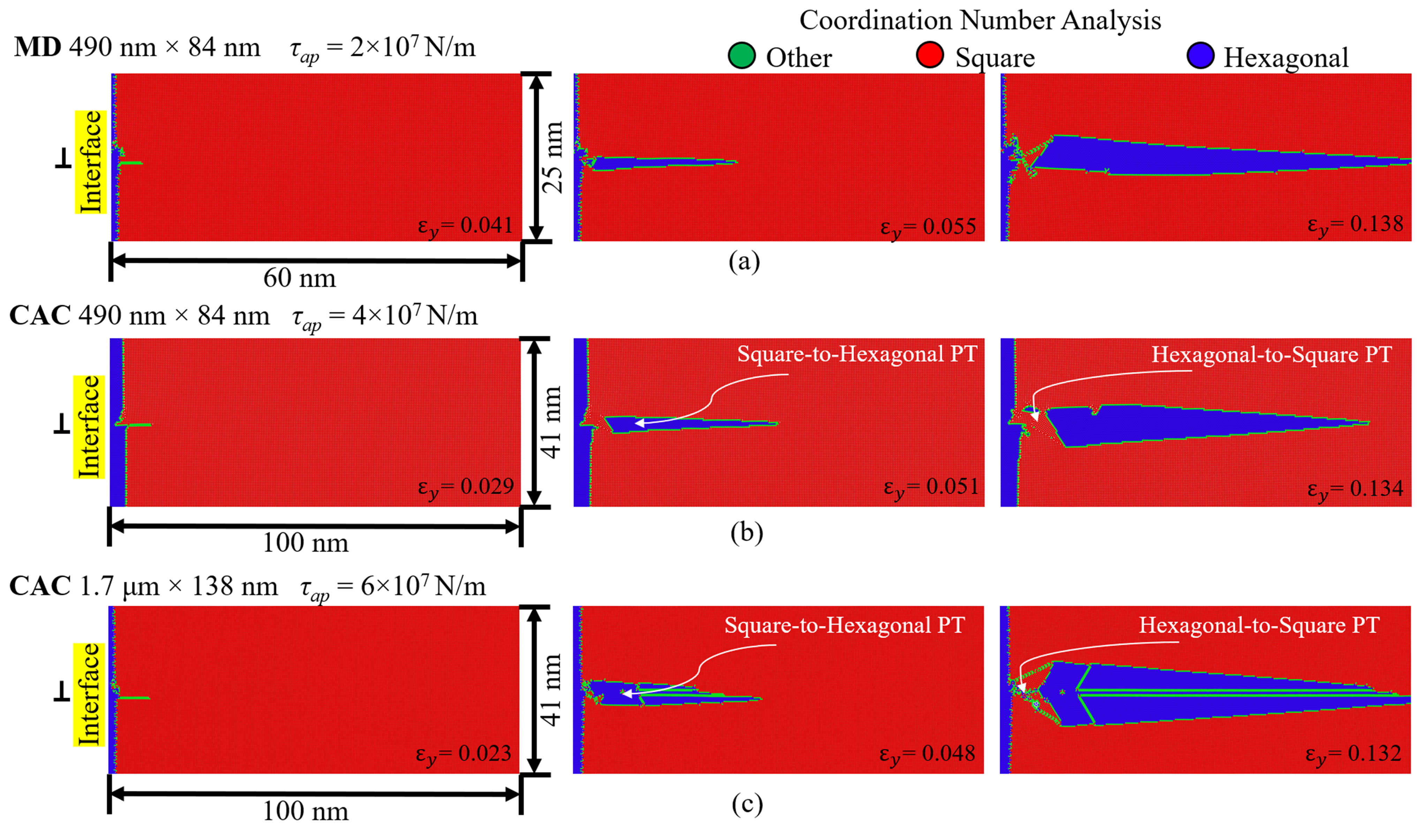}
  \caption{The snapshots of the atomic-scale structure evolution ahead of a slip-interface intersection showing the process of a square-to-hexagonal PT and then a reverse PT, i.e., the hexagonal-to-square PT, in a two-phase materials under a combined compression and shear through the coordination number analysis of the results from (a) MD simulations with $\tau_{ap}=2\times10^7$ N/m; (b) CAC simulations with $\tau_{ap}=4\times10^7$ N/m; and (c) CAC simulations with $\tau_{ap}=6\times10^7$ N/m.}
  \label{fig:structure_analysis}
\end{figure*}

\subsection{Pileup-assisted PTs and Twinning}
\subsubsection{The two-variant hexagonal phase formation}

Thereafter the dislocation pileup formation, a compression along \emph{y} direction is applied through a displacement-controlled boundary condition.
Together with the pre-shear $\tau_{ap}$, the top and bottom boundary layers of the sample are held as rigid and forced to move towards each other at 0.2 m/s.
This allows us to determine the role of a dislocation pileup in the subsequent structure changes, such as PTs, reverse PTs, and twinning, if there would be any.

\begin{figure*}
  \centering
  \includegraphics[width=0.8\paperwidth]{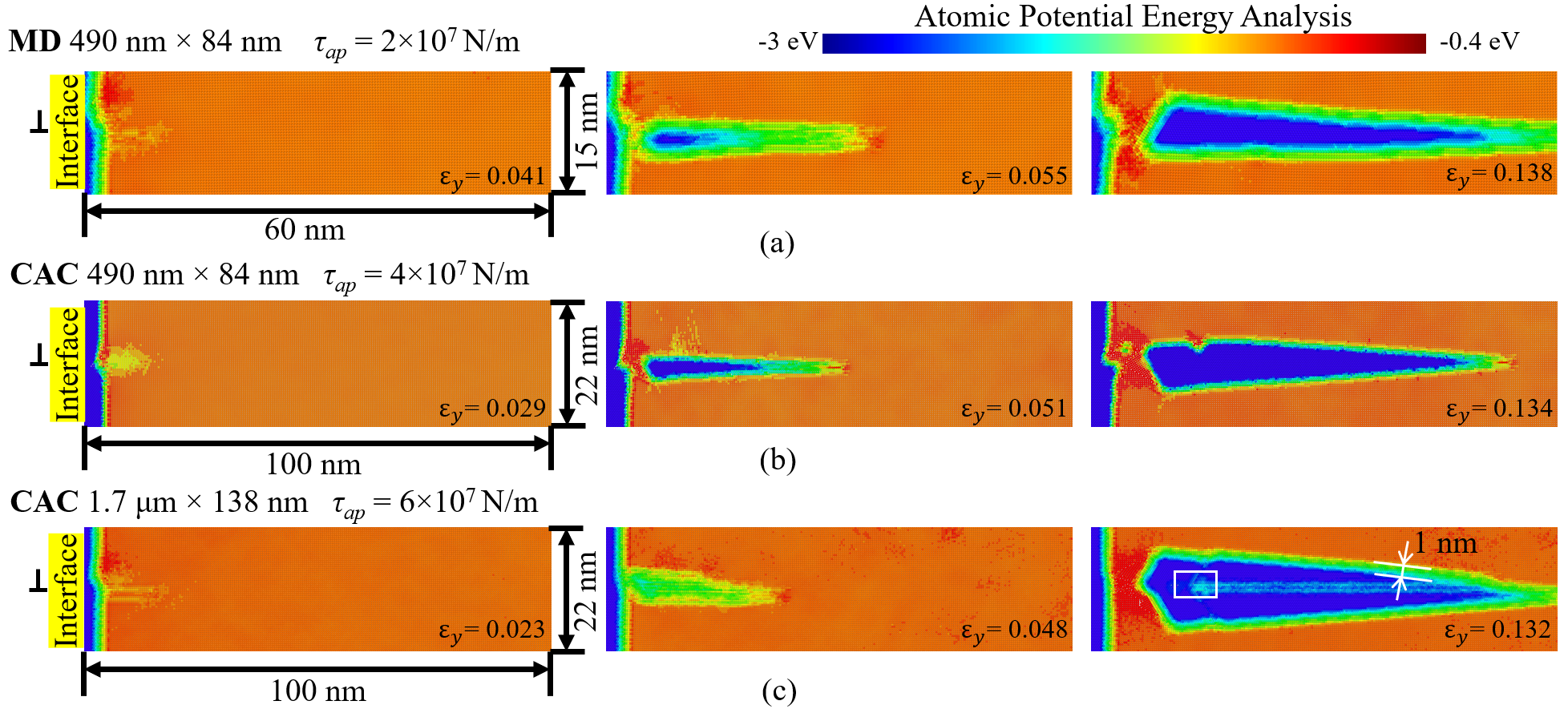}
  \caption{The snapshots of the atomic-scale structure evolution ahead of a slip-interface intersection showing the process of a square-to-hexagonal PT near a pileup tip in two-phase materials under a combined compression and shear through a potential energy analysis of the results from MD simulations with $\tau_{ap}=2\times10^7$ N/m; (b) CAC simulations with $\tau_{ap}=4\times10^7$ N/m; and (c) CAC simulations with $\tau_{ap}=6\times10^7$ N/m.}
  \label{fig:PE_analysis}
\end{figure*}

\begin{figure*}
  \centering
  \includegraphics[width=0.8\paperwidth]{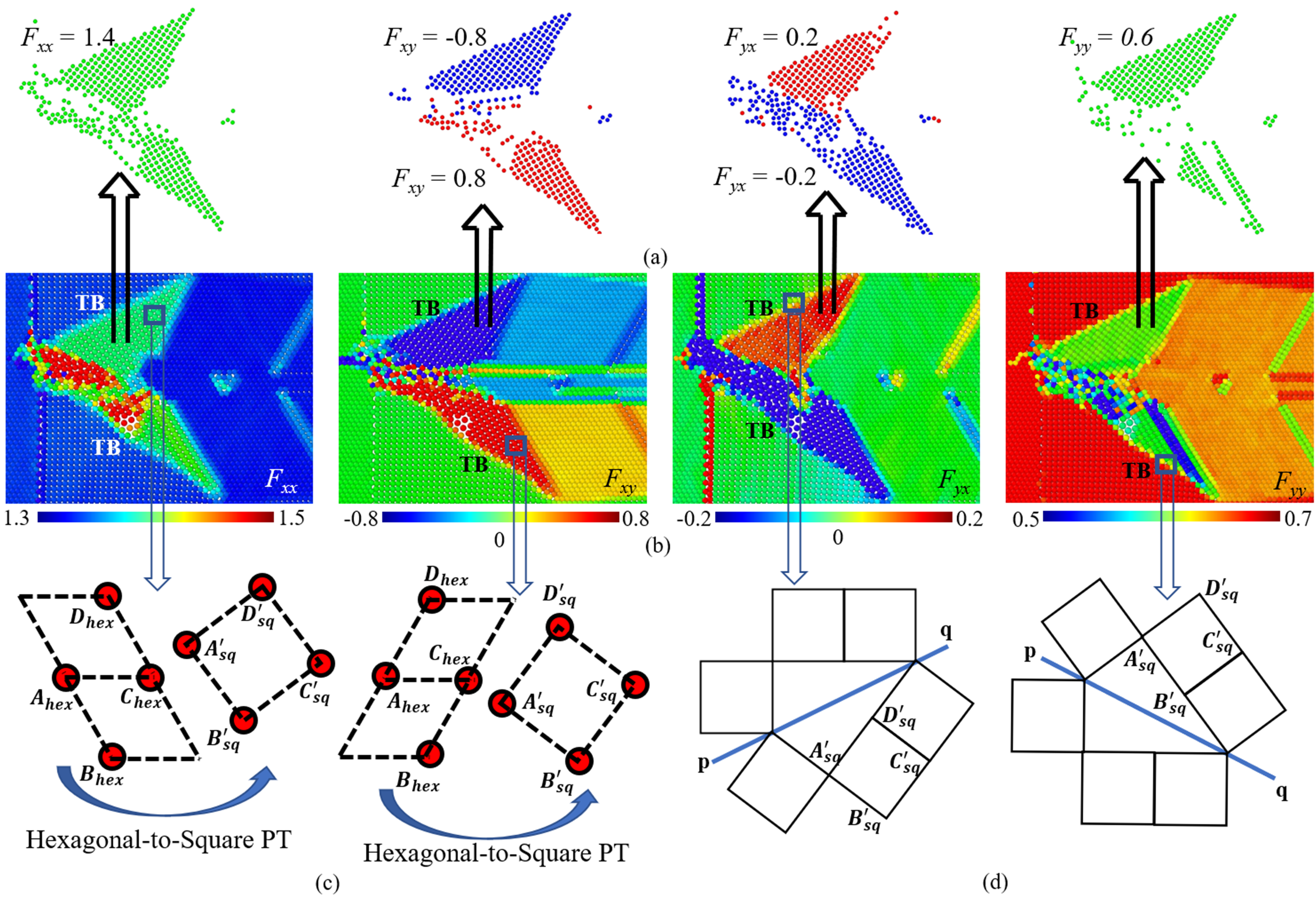}
  \caption{(a) The atomic configuration in the square phase resulting from the hexagonal-to-square PT (reverse PT).Here the atoms are colored in their deformation gradient components, $F_{xx}$, $F_{xy}$, $F_{yx}$, and $F_{yy}$, respectively. Only the atoms with $F_{xx}=1.4$, $F_{xy}=0.8$, $F_{yx}=0.2$; and $F_{yy}=0.6$ are displayed; (b) One snapshot at $t=17\ \mathrm{ps}$ showing the atomic configuration ahead of the pileup tip in CAC simulations with all the atoms being displayed and colored in $F_{xx}$, $F_{xy}$, $F_{yx}$, and $F_{yy}$, respectively; (c) an atomic-scale sketch showing how the two variants in the hexagonal phases transform back to the square phase; and (d) a sketch elucidating the formation of the twinned structure in the square phases with a TB (noted as $\boldsymbol{pq}$) in between the original square phases and the newly formed square phase resulting from the reverse PT.}
  \label{fig:square_twin}
\end{figure*}

Firstly, the dislocation pileup-induced strain localization together with the local atomic structure changes are analyzed by calculating the atomic-level deformation gradient tensor, $\boldsymbol{F}$, the Lagrangian strain tensor $\boldsymbol{E}=1/2(\boldsymbol{F}^T \cdot \boldsymbol{F}-\boldsymbol{I})$, and also the coordination number using OVITO \cite{stukowskiVisualizationAnalysisAtomistic2009}.
The time sequences of the snapshots showing the atomic structure evolution during the PT process from CAC simulation (7 dislocations have arrived at the interface and another 9 of them are behind the pileup tip) are displayed in Fig.\ \ref{fig:grain_evo}.
The atoms in the first row of Fig.\ \ref{fig:grain_evo} are colored in a shear strain component of $E_{xy}$ and the atoms in the second row are colored according to their coordination number with a cutoff of 3.92 \AA.
The coordination number analysis in Fig.\ \ref{fig:grain_evo} clearly shows that, at an early stage of the compression (from $t = 12\ \mathrm{ps}$ to $t=15\ \mathrm{ps}$), there exists a direct PT, i.e., from square (red) to hexagonal (blue), ahead of the dislocation pileup tip.
During such direct PTs (square-to-hexagonal), the PT occurs through simultaneously activating two variants of the hexagonal phase as shown in Fig.\ \ref{fig:potential}b.
When these two transformed domains are in contact with each other, a twin boundary (TB) is formed.
In details, at $t = 12\ \mathrm{ps}$, two atomic layers ahead of the pileup tip in the square phase are sheared to form a stacking fault, a precursor of the hexagonal phase.
With a further increase of the loading, the stacking faulted atomic layers transforms to the hexagonal phase at $t = 14\ \mathrm{ps}$.
The shear strain associated with the newly formed hexagonal phase above the slip plane is in an opposite sign (blue) comparing with that (orange) below this plane (see the first row of Fig.\ \ref{fig:grain_evo} from $t = 14\ \mathrm{ps}$ and $t = 17\ \mathrm{ps}$).
This corresponds to the two variants in the newly formed hexagonal phases, which are twinned with respect to each other and separated by a TB in between.
Upon a further increase of the loading, the newly formed hexagonal phase grows into a wedge shape.
The interface between the product (hexagonal) and the parent (square) phases consist of horizontal coherent regions separated by steps, avoiding the high-energy irrational interfaces.

Here the CAC simulation results, especially the material microstructure evolution ahead of the dislocation pileup, are also compared with that from MD simulations.
In details, Fig.\ \ref{fig:structure_analysis}a shows the snapshots of atomic structure evolution from MD simulations of the deformation behavior in a sample under a pre-shear of $\tau_{ap}=2\times10^7\ \mathrm{N/m}$ together with a compressive strain of $\varepsilon_{y}$ being increased from 0.054 to 0.080.
In parallel, Fig.\ \ref{fig:structure_analysis}b presents the results from a coordinate number analysis of the CAC simulations under $\tau_{ap}=4\times10^7\ \mathrm{N/m}$ together with $\varepsilon_{y}$ increasing from 0.042 to 0.063. Fig.\ \ref{fig:structure_analysis}c shows the results from CAC simulations of the dislocation pileup-assisted PT in a micron-sized sample under $\tau_{ap}=6\times10^7\ \mathrm{N/m}$ with $\varepsilon_{y}$ being increased from 0.028 to 0.039.
In Fig.\ \ref{fig:structure_analysis}a-c, the atoms in hexagonal and square phases are colored in blue and red, respectively.
The atoms in any other configuration different from hexagonal and square lattice are colored in green.
They represent either the PBs that separate the hexagonal and square phase, or the stacking faults/defects in the hexagonal phases.
Similar atomic structure evolution in the same region from MD (Fig.\ \ref{fig:structure_analysis}a) and CAC (Fig.\ \ref{fig:structure_analysis}b and c) simulations is also analyzed in terms of atomic potential energy (PE) as shown in Fig.\ \ref{fig:PE_analysis}, where the atoms in hexagonal and square phases are colored in blue and orange, respectively, while those atoms with PE values deviating from the hexagonal or square lattice are in green.

Our major observations from Fig.\ \ref{fig:structure_analysis} and Fig.\ \ref{fig:PE_analysis} are:
\textbf{(1)} the square-to-hexagonal PT occurs at $\varepsilon_{y}$ = 0.028 when $\tau_{ap}=6\times10^7\ \mathrm{N/m}$ is applied on a micron-sized CAC model where tens of dislocations participate the pileup, while $\varepsilon_{y}$ = 0.054 when $\tau_{ap} = 2\times10^7\ \mathrm{N/m}$ is applied on nanoscale MD models where only a few dislocations participate the pileup;
\textbf{(2)} in both CAC and MD simulations, the PT starts at the dislocation pileup tip.
The newly nucleated hexagonal phase then grows into a wedge shape.
Such a wedge-shape in the product phase resulting from a dislocation-assisted PT has also been observed in experiments \cite{ibarraMartensiteNucleationDislocations2007}, although in a different material;
\textbf{(3)} the width of the interface between the parent and the product phases can be approximately estimated to be about 1 nm in both CAC and MD simulations;
This value can be used for calibrating the PFA models in \cite{levitasPhaseTransformationsNanograin2014, javanbakht2016phase, javanbakht2018nanoscale};
\textbf{(4)} the stacking fault in the product hexagonal phase only appears in the microscale CAC simulations at $\tau_{ap}=6\times10^7$ N/m, suggesting an increased microstructure complexity when more dislocations are piled up.

\subsubsection{Twinning in square phase via a reverse PT}

Interestingly, at a later stage of the deformation, the coordination number analysis in the second row of Fig.\ \ref{fig:grain_evo} when $t =15\ \mathrm{ps}$ and $t=17\ \mathrm{ps}$ shows that, at the slip-interface intersection, the newly formed hexagonal phases ahead of the pileup tip transform back to the square phases.
This is referred as a reverse PT, i.e., hexagonal-to-square PT.
The finish of the square-to-hexagonal PT and then the onset of a reverse PT result from the large back stresses due to transformation shear and an atomic structure reconfiguration at the slip-interface intersection, which has largely released the local internal stresses.
The newly formed square phase resulting from such a reverse PT together with the original square phases form a twinned structure.
Although the twinning formation through a direct PT was reported in literature
\cite{bhattacharya2004crystal, castany2016reversion, ombogo2020nucleation, zahiri2021formation}, our finding here is believed to be the first direct prediction of twinning formation during a monotonous loading through a dislocation pileup-induced PT and then a reverse PT.
In other words, twinning does not occur directly in the square phase due to a high energy barrier there, but occurs via a direct-to-reverse PT instead.
Analytically, when the elastic strains are neglected, the transformation deformation gradients, $\boldsymbol{U}_i$ and $\boldsymbol{U}_j$, in the two regions across a twin boundary (TB) satisfy

\begin{figure}
  \centering
  \includegraphics[width=0.4\paperwidth]{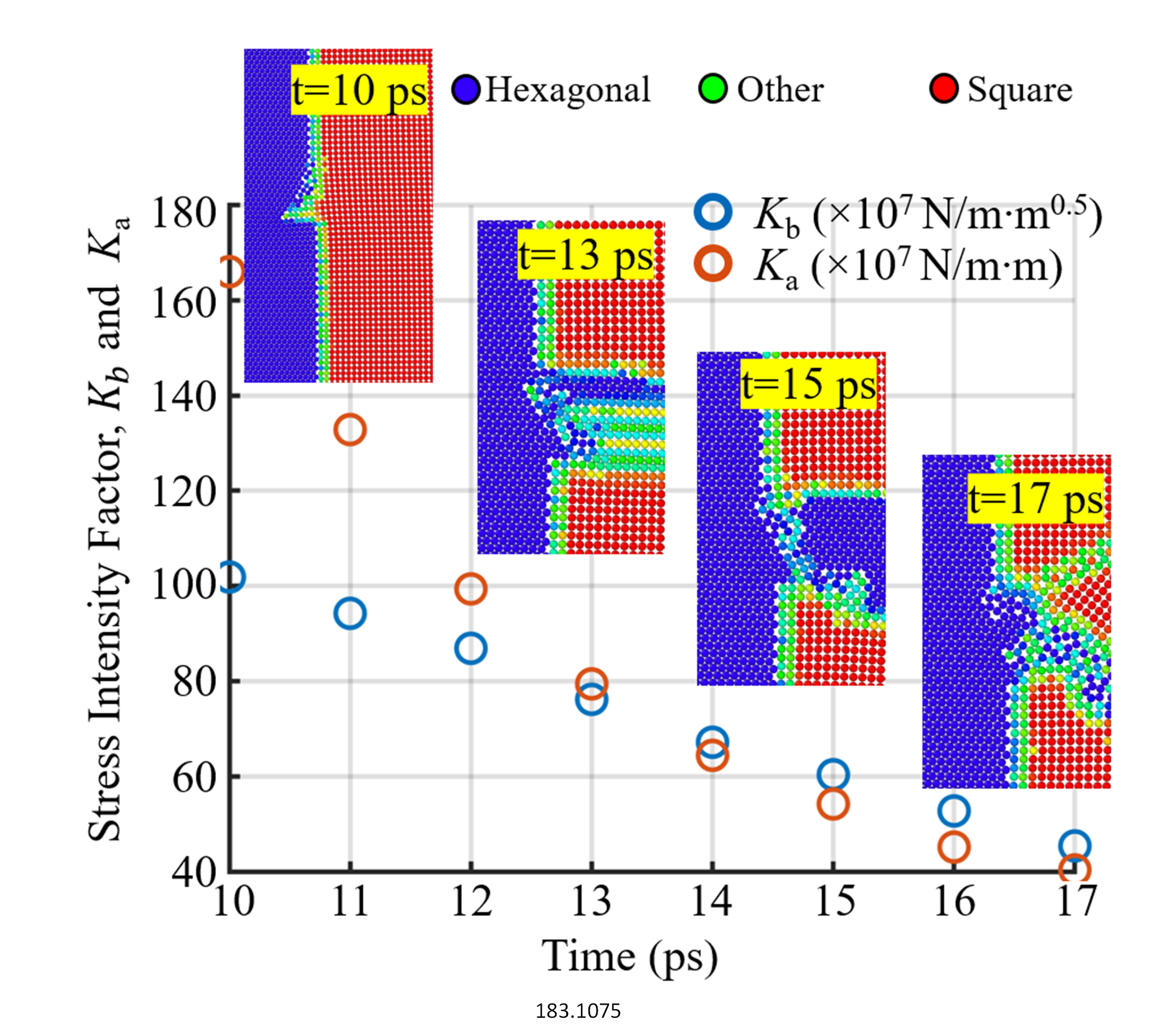}
  \caption{The time evolution of the stress intensity factors, $K_b$ and $K_a$, ahead of the dislocation pileup tip with the inset pictures showing the snapshots of the local atomic structure changes during the direct and then reverse PTs.}
  \label{fig:kevo}
\end{figure}

\begin{figure*}
  \centering
  \includegraphics[width=0.8\paperwidth]{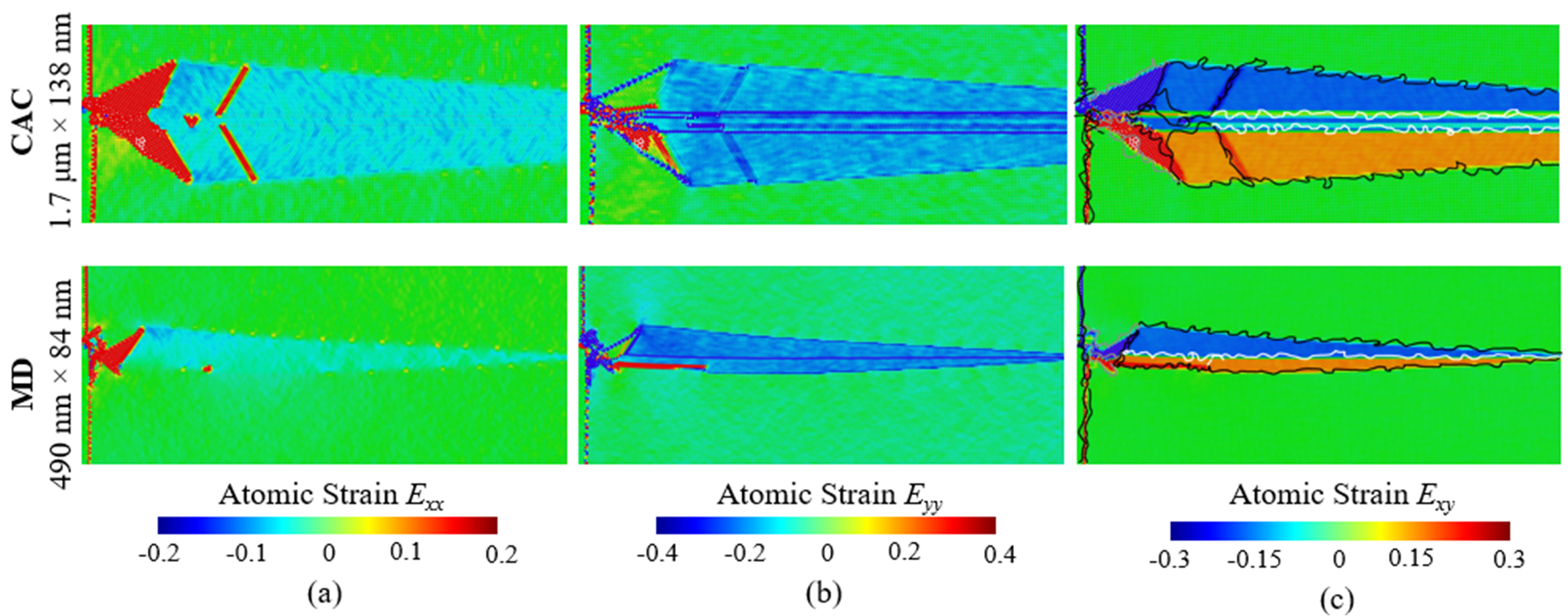}
  \caption{The contour of the atomic-level strain at $t=17\ \mathrm{ps}$: (a) $E_{xx}$; (b) $E_{yy}$; and (c) $E_{xy}$ showing the simultaneous occurrence of twinning and PTs ahead of the dislocation pileup from CAC and MD simulations. This solution corresponds to the stationary state. The black lines correspond to the phase interface equilibrium condition of $X=0$ with deformation gradients being taken from the parent and product phases, while the white lines correspond to the twin interface in an equilibrium condition of $X=0$ with the deformation gradients being taken from the twinning variants across the TB.}
  \label{fig:twin_PTwork}
\end{figure*}

\begin{equation}
  \boldsymbol{Q}_i\cdot\boldsymbol{U}_{ti}-\boldsymbol{Q}_j\cdot\boldsymbol{U}_{tj}=\gamma_t \boldsymbol{m}\boldsymbol{n}.
  \label{eq:twinning_equation}
\end{equation}

\noindent Here, $\gamma_t$ is the twinning shear,
$\boldsymbol{m}$ is the twinning direction within the twinning plane which has a unit normal of $\boldsymbol{n}$.
Eq.\ (\ref{eq:twinning_equation}) is used here to check whether the newly formed square phase resulting from the reverse PT is indeed twinned with respect to the original square phase.
For the observed reverse PT, $i$ represents an original square phase, and $j$ is for the domain in which the material is in a square phase resulting from the hexagonal-to-square PT.
Then $ \boldsymbol{Q}_i=\boldsymbol{U}_i=\boldsymbol{U}_j= \boldsymbol{I}$, and Eq.\ (\ref{eq:twinning_equation}) is simplified as

\begin{equation}
  \boldsymbol{I}-\boldsymbol{Q}=\gamma_t \boldsymbol{m}\boldsymbol{n},
  \label{eq:twinning_equation-1}
\end{equation}
\noindent which has at least two solutions as following:

\begin{eqnarray}
  \gamma^+ = 0.25; \qquad   \boldsymbol{m}^+=\{4/\sqrt{5}, 2/\sqrt{5}\}^T; \qquad
  \nonumber \\  \boldsymbol{n}^+=\{2/\sqrt{5}, -4/\sqrt{5}\};
  \qquad
  \boldsymbol{Q}^+= \left(\begin{matrix}  1.4 &  -0.8 \\  0.2 & 0.6 \end{matrix}\right);
  \nonumber \\
  \label{eq:Fsquare1}
\end{eqnarray}

or

\begin{eqnarray}
  \gamma^- = 0.25;   \qquad
  \boldsymbol{m}^-=\{4/\sqrt{5}, -2/\sqrt{5}\}^T; \qquad
  \nonumber \\  \boldsymbol{n}^-=\{2/\sqrt{5}, 4/\sqrt{5}\};
  \qquad
  \boldsymbol{Q}^-= \left(\begin{matrix}  1.4 &  0.8 \\  -0.2 & 0.6 \end{matrix}\right).
  \label{eq:Fsquare2}
\end{eqnarray}

\noindent To compare our simulation results against the above analytical solutions, the atomic-level deformation gradient components, $F_{xx}$, $F_{xy}$, $F_{yx}$, and $F_{yy}$, from CAC simulations at $t$=17 ps are calculated using the initially un-deformed atomic configuration as a reference and shown in Fig.\ \ref{fig:square_twin}a.
It is seen that the simulation-based $\boldsymbol{F}$ in the region where the reverse PT occurs agrees with the above analytical solution of $\boldsymbol{Q}$ very well.
Therefore, the square phase resulting from such a reverse PT is indeed twinned with respect to the original square phase.
In more details, for the region where the reverse PT occurs, the atomic configuration at the dislocation pileup tip is shown in Fig.\ \ref{fig:square_twin}b.
Here all the atoms are displayed and colored by their deformation gradient, $F_{xx}$, $F_{xy}$, $F_{yx}$, and $F_{yy}$, respectively.
Based on the detailed crystallographic orientation analysis (Fig.\ \ref{fig:square_twin}c), the TB formation process is also sketched in Fig.\ \ref{fig:square_twin}d for clarifications.

During compression, accompanied by the direct and reverse PTs, the local stresses ahead of the dislocation pileup evolves.
To characterize the full complexity of such a complex internal stress evolution, we measure the local stresses and again decompose it into two parts:

\begin{equation}
  \tau_{xy}=\frac{K_b}{\sqrt{\pi(r+r_0)}}+\frac{K_a}{(r+r_0)}+\tau_0,
  \label{eq:stressdistribution_kafitting}
\end{equation}

\noindent in which $K_b$ and $K_a$ are noted as the intensity factors based on the Eshelby and super-dislocation model, respectively.
During the process of the direct and reverse PT, a series of stress profiles ahead of the pileup tip are produced.
A fitting of each stress profile into the above equation leads to a determination of the instantaneous $K_b$ and $K_a$.
A time evolution of $K_b$ and $K_a$ together with the snapshots of the atomic structure at the pileup tip are presented in Fig.\ \ref{fig:kevo}.
Obviously, before the occurrence of PT, both $K_b$ and $K_a$ are at a relatively high level.
And the stress intensity, $K_a$, induced by the dislocation at the pileup tip is larger than that, $K_b$.
Both $K_b$ and $K_a$ decrease when PTs occur.
In details, at $t=13$ ps, with the growth of the hexagonal phase, $K_a$ decreases from 165 to 80 while $K_b$ decreases from 100 to 75 (all in a unit of $10^7\, N/m^{1/2}$).
It suggests that a direct PT largely relaxes the local stress concentration.
Thereafter, the step at the slip-interface intersection is found to disassociate into multiple mini-steps.
At this stage, the reverse PT starts and those mini-steps eventually emerge as a single step at $t=15$ ps.
Up to this moment, the stress intensity becomes low.
At $t=14$ ps, there is a cross-over of the stress intensity induced by the dislocations behind the pileup tip: $K_b$ becomes larger than that, $K_a$, induced by the dislocations at the pileup tip.
\bigskip

\subsubsection{Thermodynamic driving force at the interfaces}
In addition to the atomic-level deformation gradient tensor as shown in Fig.\ \ref{fig:square_twin}, for the snapshot at $t=17\ \mathrm{ps}$ in Fig.\ \ref{fig:grain_evo}, three components of the Lagrange strain tensor, $E_{xx}$, $E_{yy}$ and $E_{xy}$, associated with each atom ahead of the slip-interface intersection are also extracted and presented in Fig.\ \ref{fig:twin_PTwork}.
In order to quantify how the local stress contributes to the subsequent PTs, an atomic-level local strain together with a local stress-based thermodynamic driving force analysis (similar to that for nanoscale \cite{javanbakht2016phase, javanbakht2018nanoscale} and scale-free \cite{levitasScalefreeModelingCoupled2018, esfahaniStraininducedMultivariantMartensitic2020, babaei2020finite} PFAs) are also performed here.
Based on the atomic-level local stresses, the thermodynamic Eshelby-type driving force for PT and twinning, $X$ is defined as:

\begin{equation}
  X=\textbf{\emph{P}}^T:(\textbf{\emph{F}}_{t2}-\textbf{\emph{F}}_{t1})-\Delta\Psi.
  \label{eq:PTWork}
\end{equation}

\noindent Here, $\boldsymbol{P} = J \boldsymbol{\sigma} \cdot \boldsymbol{F}^{-1T} $
is the first Piola-Kirchhoff stress expressed in terms of the Cauchy stress $\boldsymbol{\sigma}$ with $ \emph{J} =det \textbf{\emph{F}}$.
$\Delta\Psi$ is potential energy difference between the stress-free phases from both sides of each interface (which is zero for twin boundaries).
$\boldsymbol{F}_{t1}$ and $\boldsymbol{F}_{t2}$ are the transformation deformation gradients in the two material domains across an interface.
In particular, for the TBs in the hexagonal phases, they are $\boldsymbol{F}_t^+$ and $\boldsymbol{F}_t^-$ in Eq.\ (\ref{eq:F}), respectively.
For the TBs in the square phases ahead of the pileup tip, they are $\boldsymbol{I}$ and $\boldsymbol{Q}^+$ or $\boldsymbol{Q}^+$ in Eq.\ (\ref{eq:Fsquare1}) and Eq.\ (\ref{eq:Fsquare2}).
For the PBs between hexagonal and square phases, they are either $\boldsymbol{F}_t^+$ or $\boldsymbol{F}_t^-$ and the transformation deformation gradient in the square phase, $\boldsymbol{I}$, $\boldsymbol{Q}^+$ or $\boldsymbol{Q}^-$.
The condition of $X = 0$ corresponds to the local thermodynamic equilibrium of the relevant interface under consideration.

The black lines in Fig.\ \ref{fig:twin_PTwork}c correspond to the phase interface equilibrium condition of $X=0$ with the deformation gradients in it being taken from the parent and product phases across the interface.
In contrast, the white lines correspond to the twin interface equilibrium condition of $X=0$ with the deformation gradients in it being taken from the two twinning variants across the TB.
Figure\ \ref{fig:twin_PTwork}c shows that, during the processes of direct PTs, reverse PTs, and twinning, all interfaces (including initial vertical phase interface) correspond to the continuum phase equilibrium condition of $X=0$.
Thus, these interfaces are stationary.
Although this has been shown for cubic-tetragonal phase interface in nanoscale \cite{javanbakht2016phase, javanbakht2018nanoscale} and scale-free \cite{levitasScalefreeModelingCoupled2018, esfahaniStraininducedMultivariantMartensitic2020, babaei2020finite} PFAs at the continuum level, it is observed for the first time here at the atomic scale with a variety of different twinning interfaces being included.

\subsection{The Role of the Pileup-Induced Local Stresses in the Subsequent PTs}

\begin{figure}
  \centering
  \includegraphics[width=0.4\paperwidth]{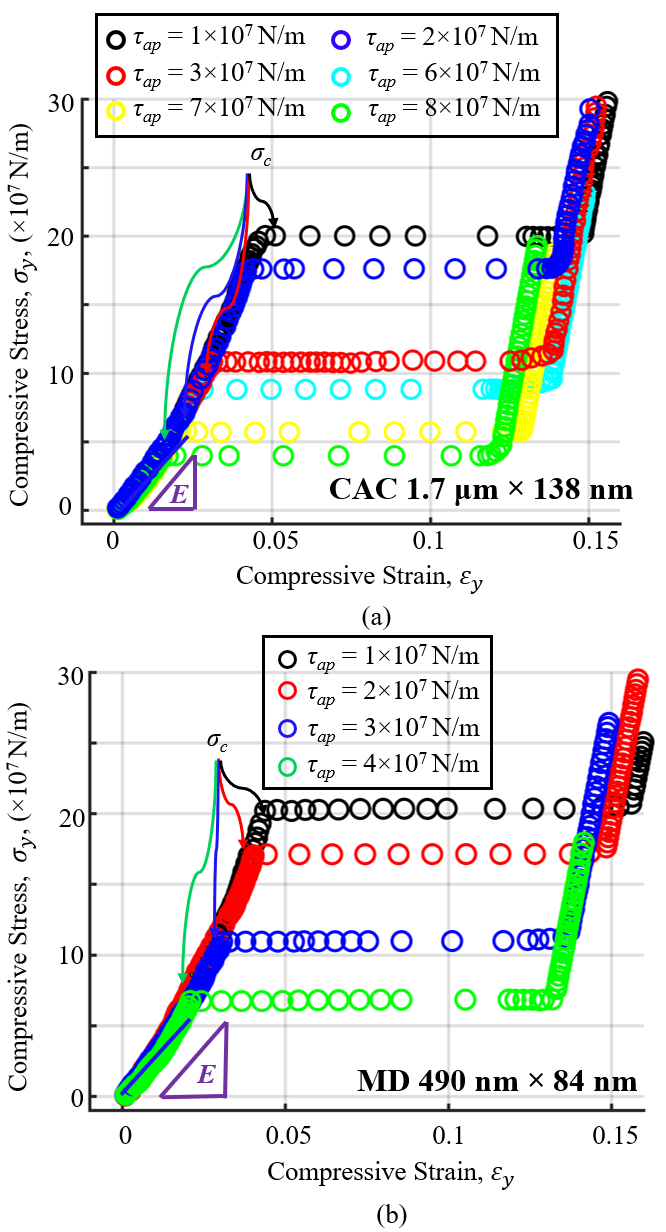}
  \caption{The stress-strain curves from (a) CAC and (b) MD simulations of PTs in multi-phase lattices under compression after the pileup of a certain number of dislocations at the interface under different $\tau_{ap}$.}
  \label{fig:stress_strain}
\end{figure}

\begin{figure}
  \centering
  \includegraphics[width=0.4\paperwidth]{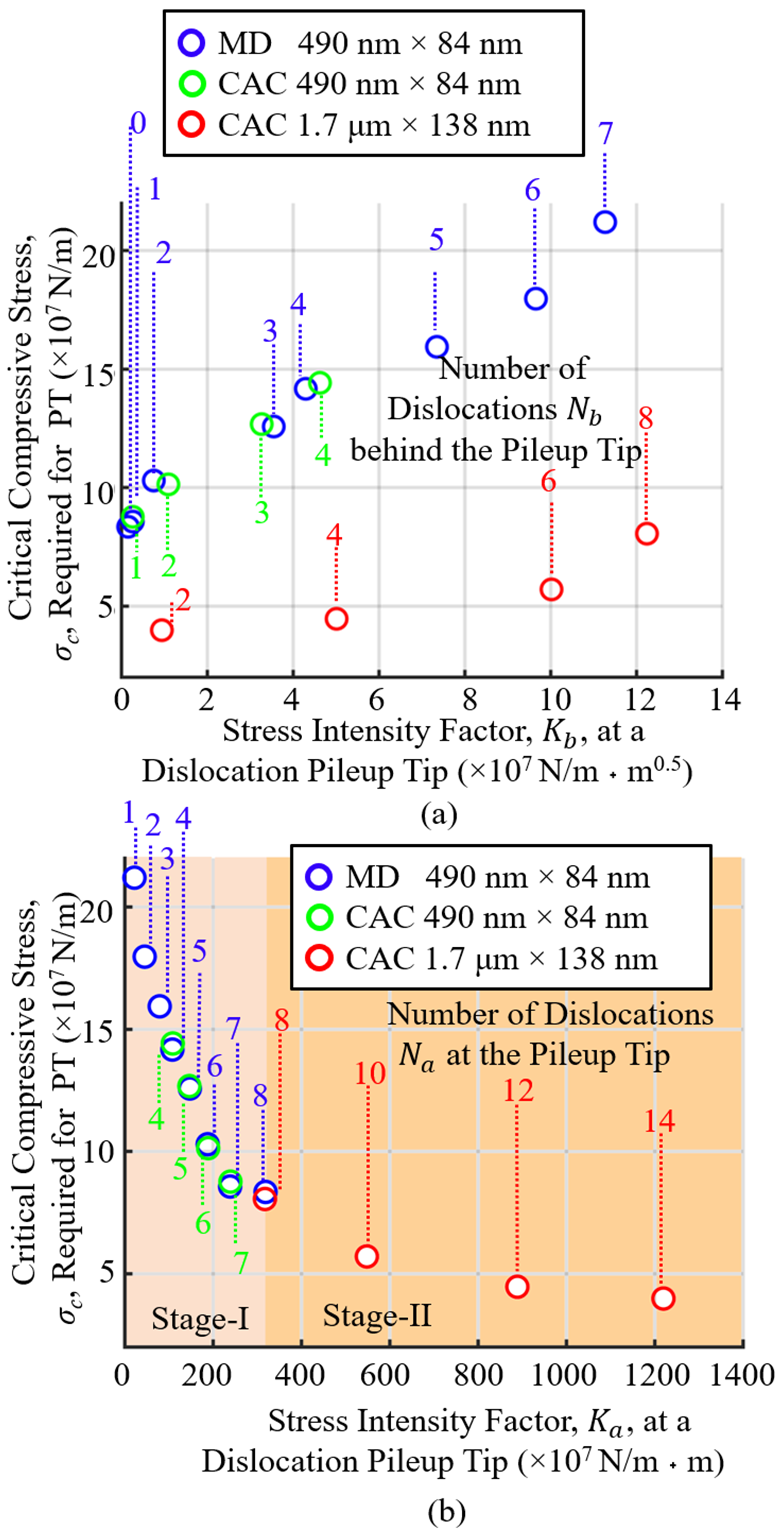}
  \caption{The relationship between the stress intensity factors (a) $K_b$ behind the pileup tip, (b) $K_a$ ahead of the pileup tip and the critical compressive stress $\sigma_c$ required for activating the PT in both the nanoscale MD and microscale CAC simulations.}
  \label{fig:stress_reduction}
\end{figure}

In order to quantify how the dislocation pileup-induced local stresses contribute to the subsequent PT, we perform a series of analysis on how the square phase responds to compression through plotting the ($\sigma_{y}$- $\varepsilon_{y}$) curves (Figs.\ \ref{fig:stress_strain}a-b).
Here, $\varepsilon_{y}$ is the Lagrangian strain along $y$-direction.
The stress, $\sigma_{y}$, is an average of the true stresses along $y$-direction acting on the atoms located in a domain ($100\ \mathrm{nm}\times22\ \mathrm{nm}$ in CAC and $60\ \mathrm{nm}\times15\ \mathrm{nm}$ in MD, respectively) ahead of the dislocation pileup tip.
Results in Figs.\ \ref{fig:stress_strain}a-b show that:
\textbf{(i)} prior to $\varepsilon_{y}$ = 0.02, both CAC and MD simulations predict a linear stress-strain relation with the same modulus of \emph{E}.
At larger strains, the material behaviour is highly nonlinear though;
\textbf{(ii)} thereafter, a plateau appears on all the stress-strain curves;
\textbf{(iii)} a square-to-hexagonal PT is found to start at $\varepsilon_{y}$ where the plateau appears.
This corresponds well to the atomic configuration evolution in Fig.\ \ref{fig:structure_analysis} and Fig.\ \ref{fig:PE_analysis};
\textbf{(iv)} the critical compressive stress required for the occurrence of PT can be thus identified as $\sigma_c$ in Figs.\ \ref{fig:stress_strain}a-b; and
\textbf{(v)} $\sigma_c$ largely decreases when the pre-applied shear is increased from $\tau_{ap} = 2\times10^7\ \mathrm{N/m}$ to $\tau_{ap} = 8\times10^7\ \mathrm{N/m}$.
The considerable reduction of $\sigma_c$ with the increase of $\tau_{ap}$, i.e., from $\sigma_c = 16\times10^7\ \mathrm{N/m}$ at $\tau_{ap} = 2\times10^7\ \mathrm{N/m}$ to $\sigma_c = 4\times10^7\ \mathrm{N/m}$ at $\tau_{ap} = 8\times10^7\ \mathrm{N/m}$, can be attributed to the large number of dislocations accumulated at the interface, which has introduced a high local stress assisting the PT at the pileup tip.

Furthermore, Fig.\ \ref{fig:stress_reduction} presents the data from both CAC and MD simulations for quantitatively relating the critical stress, $\sigma_c$, at which the PT starts, with the previously determined stress intensity factor $K_b$ and $K_a$.
Fig.\ \ref{fig:stress_reduction} clearly shows:
\textbf{(i)} the $\sigma_c$ required for initiating the PTs can be as high as $22\times10^7\ \mathrm{N/m}$ if only one dislocation has been initially introduced into the model, but can be reduced to $4\times10^7\ \mathrm{N/m}$ when 16 dislocations are piled up at the interface.
This corresponds to PT pressure reduction by a factor of 5.5; and
\textbf{(ii)} the $\sigma_c-K_a$ relation can be divided into to two stages.
Upon an increase of $K_b$ and $K_a$, $\sigma_c$ in Stage-I reduces considerably faster than it does in Stage-II.
It should be also noted that, for the material system under consideration here, $\sigma_y$ is not the only stress component that contributes to the PT.
Our recent simulations show that all the stress components, i.e., $\sigma_x$, $\sigma_y$, and $\tau_{xy}$, all contribute to PTs, although the magnitude of $\sigma_x$ in the present simulations is found to be considerably small.
A crystal lattice instability criterion based on all components of the stress tensor will be reported in a separate paper.
\bigskip

\section{CONCLUDING REMARKS}
\label{conclusion}

To summarize, here we present an atomistic-to-microscale computational analysis of the interplay between dislocation slip and PT/twinning in two-phase materials under compression and shear.
One main novelty of the CAC approach deployed here is its capability in bridging the relevant length scales by resolving the structure changes near an interface at the atomic scale while the lagging dislocations away from the interface in a coarse-grained atomistic description.
It thus expands the MD-simulation-based predictive capability from the nanoscale to the micrometer level.
The main findings of this study are summarized as below:

\textbf{(1)} The micron-sized CAC model can accommodate up to 16 dislocations in one slip at a modest computational cost.
These dislocations may be blocked by the obstacles (an incoherent interface in this work) and form a pileup spanning a range of several micrometers (1.2 $\mu$m in the present model).
In contrast, the nanoscale MD model using the same computational resource can only accommodate up to 8 dislocations in a pileup, the equilibrium configuration of which under certain shear stress only spans tens of nanometers;

\textbf{(2)} The internal stress intensity induced by a dislocation pileup can be decomposed into two parts.
One (noted as $K_b$) follows the Eshelby model and is contributed by the dislocations behind the pileup tip, the other (noted as $K_a$) follows the super-dislocation model and is caused by the step formed at the slip-interface intersection.
Either the Eshelby or the super-dislocation model alone will largely underestimate the dislocation pileup-induced internal stress concentration.
The local internal stress intensity is linearly proportional to the number of the dislocations in a pileup at the nanoscale, but upper bends to a very high level when tens of dislocations participate the pileup at the microscale.

\textbf{(3)} When the pre-sheared material sample is subjected to a compression, PTs and twinning ahead of the dislocation pileup where the local stresses concentrate occur through a two-step process:
(a) the square phase transforms to the hexagonal phase.
During such a square-to-hexagonal PT (also referred as the direct PT), two variants of the hexagonal phase simultaneously nucleate, grow, contact, and form a twin boundary in between;
(b) at a later stage, a portion of the newly formed hexagonal phase transforms back to the square phases.The newly formed square phase resulting from this reverse PT is found to be twinned with respect to the original square lattices and relax the local internal stresses.
This is the first direct observation of twinning formation from a dislocation slip-assisted PTs and reverse PTs during monotonous loading.

\textbf{(4)} The Eshelby driving forces at all the newly formed PBs or TBs are zero, confirming that all of them are in the thermodynamic equilibrium.
Although such analysis was performed previously in the nano- and microscale PFAs at the continuum level, it was performed for the first time for atomistic modeling.
This can be an evidence of the consistency between the atomistic and continuum thermodynamic treatments in our model.

\textbf{(5)} The direct PT largely relaxes the local stress concentration due to the transformation strain.
Thereafter, the step at the slip-interface intersection is found to disassociate into multiple mini-steps.
The reverse PT then starts and the mini-steps eventually emerge as one step.

\textbf{(6)} The critical compressive stress, $\sigma_c$, required for initiating the PT, largely decreases with the increase of the internal stress intensity factor.
Our simulations suggest a possibility of reducing $\sigma_c$ for PT by a factor of 5.5 and even higher through injecting a considerably large number of dislocations into a micron-sized material sample, as observed in experiments for various material systems.
\bigskip

These findings highlight:
\textbf{(a)} the insufficiency of only using nanoscale MD simulations to interpret the experimental observations on the slip-interface reaction, which may have involved hundreds of $\mu$m-long dislocations participating in the formation of a pileup; and
\textbf{(b)} the possibility of using CAC to predict how the microscale dislocation-mediated plastic flow reacts with the buried interfaces in a variety of multi-phase materials, such as fcc/bcc, fcc/hcp, bcc/hcp metallic composites, Ti-/Zr-/high entropy alloys, among others, when subjected to a severe deformation.
In such scenarios, the CAC simulation tool may provide researchers with suitable, if not the best, vehicle for simultanesouly considering the microscale plasticity together with the atomic-scale interface structure relaxation.
Nevertheless, this framework is considered to be still at a preliminary stage because:

\textbf{(i)} Comparing with MD, the length scale of the present CAC models is indeed one step closer but is not the same level as that in experiments yet.
In particular, as far as the phase growth after the slip-interface reactions are concerned, a further length scaling up to microns through a CG description of the PTs in the material domain far away from the interface is needed.
Otherwise, the growth of new phases from CAC simulations is confined at the nanoscale;
\textbf{(ii)} In addition to the local stress, another factor that plays a vital role in the PT process is the thermal-induced atomic fluctuations.
The implementation of a finite temperature algorithm into CAC for capturing the finite-temperature effects on dislocations, PTs, and their interactions is needed, especially when the correlation between phonon instability and PTs becomes a concern;
\textbf{(iii)} The material system under consideration here is oversimplified in terms of crystal structure, chemistry, interatomic potential, and microstructure.
A transfer of the present model or the gained knowledge for understanding the dislocations-interface reactions and subsequent structural changes in realistic multi-phase materials is not trivial.
It demands:
(1) the design of new finite elements for simultaneously accommodating complex dislocation activities;
(2) the implementation of more sophisticated or machine learning-based interatomic potential to be trained from \emph{ab initio} data for capturing more phase variants, as well as
(3) the incorporation of realistic structures and chemistry at the GBs, PBs, and other interfaces in the materials.
An expansion of CAC along those three directions and its applications in predicting the slip-interface reactions in real materials are being attempted and will be reported in the future.
\bigskip

\section*{Acknowledgments}
YP, RJ, TP, and LX acknowledge the support of the U.S. National Science Foundation (CMMI-1824840 and CMMI-1930093) and the Extreme Science \& Engineering Discovery Environment (XSEDE-TG-MSS170003, XSEDE-TG-MSS190008, and XSEDE-TG-MSS190013). LC acknowledge support from the U.S. Dept. of Energy, Office of Basic Energy Sciences Project FWP 06SCPE401. YP and LC also acknowledge the support of NSF and Los Alamos National Laboratory through a NSF INTERNSHIP program under DMR-1807545. VIL work was funded by NSF (MMN-1904830 and CMMI-1943710) and the ISU (Vance Coffman Faculty Chair Professorship), as well as XSEDE, allocation TG-MSS170015.

\bibliography{ref}

\end{document}